\documentclass[traditabstract]{aa}
\pdfoutput=1
\usepackage{graphicx}
\usepackage{txfonts}
\usepackage{longtable}
\usepackage[authoryear]{natbib}
\bibpunct{(}{)}{;}{a}{}{,}

\def\ltsima{$\; \buildrel < \over \sim \;$}
\def\simlt{\lower.5ex\hbox{\ltsima}}
\def\gtsima{$\; \buildrel > \over \sim \;$}
\def\simgt{\lower.5ex\hbox{\gtsima}}
\newcommand{\HI}{\ion{H}{i}}
\newcommand{\HII}{\ion{H}{ii}}
\begin{document}
   \title{An optical and \HI\  study of the dwarf Local Group galaxy VV124=UGC4879.}
\subtitle{A gas-poor dwarf with a stellar disk?}

   \author{M. Bellazzini\inst{1}, G. Beccari\inst{2}, T.A. Oosterloo\inst{3,4}, 
           S. Galleti\inst{1}, A. Sollima\inst{5,9}, M. Correnti\inst{1},\\
           V. Testa\inst{6}, L. Mayer\inst{7,8}, M. Cignoni\inst{1,10}, 
           F. Fraternali\inst{10},\and S. Gallozzi\inst{6}
           }
         
      \offprints{M. Bellazzini}

   \institute{INAF - Osservatorio Astronomico di Bologna,
              Via Ranzani 1, 40127 Bologna, Italy\\
            \email{michele.bellazzini@oabo.inaf.it} 
            \and
            European Southern Observatory, Karl-Schwarzschild-Str. 2, D-85748, 
            Garching bei Munchen, Germany
            \and
            Netherlands Institute for Radio Astronomy, Postbus 2, 7990 AA Dwingeloo, the Netherlands
            \and
            Kapteyn Astronomical Institute, University of Groningen, Postbus 800, 9700 AV Groningen, 
            The Netherlands
            \and
            Instituto de Astrofisica de Canarias, c/via Lactea s/n, La Laguna 38205-E, Spain
            \and
             INAF - Osservatorio Astronomico di Roma, via Frascati 33, 00040 Monteporzio, Italy
             \and
            Institute for Theoretical Physics, University of Z\"urich, 
            Winterthurer-strasse 190, CH-9057, Z\"urich, Switzerland
            \and 
            Institut f\"ur Astronomie, ETH Z\"urich-H\"onggerberg, 
            Wolfgang-Pauli-Strasse16, CH-8093 Z\"urich, Switzerland
            \and
            INAF - Osservatorio Astronomico di Padova, vicolo dell'Osservatorio 5, 35122, Padova, Italy
            \and
           Dipartimento di Astronomia - Universit\`a degli Studi di Bologna,
            Via Ranzani 1, 40127 Bologna, Italy\\
           }

     \authorrunning{M. Bellazzini et al.}
   \titlerunning{An optical and \HI\  study of the dwarf galaxy VV124=UGC4879.}

   \date{Accepted for publication in {\em Astronomy \& Astrophysics} }

\abstract{
We present a detailed study of the dwarf galaxy VV124 (UGC4879), recently recognized as a remarkably isolated member of the Local Group. We have obtained deep (r$\simeq 26.5$) wide-field ($23\arcmin \times 23\arcmin$) g,r photometry of individual stars with the LBC camera at the Large Binocular Telescope under sub-arcsec seeing conditions,. The Color-Magnitude Diagram suggests that the stellar content of the galaxy is dominated by an old,  metal-poor population, with a significant metallicity spread. A very clean detection of the RGB tip allows us to derive an accurate distance of  $D=1.3\pm 0.1$~Mpc.
Combining surface photometry with star counts, we are able to trace the surface brightness profile of VV124 out to $\sim 5\arcmin \simeq 1.9$~kpc radius (where $\mu_r\simeq 30$~mag/arcsec$^2$), showing that it is much more extended than previously believed. Moreover, the surface density map reveals the presence of two symmetric flattened wings emanating from the central elongated spheroid and aligned with its major axis, resembling a stellar disk seen nearly edge-on. We also present \HI\  observations obtained with the Westerbork Synthesis Radio Telescope (WSRT), the first ever of this object.
A total amount of $\simeq 10^6~M_{\sun}$ of \HI\  gas is detected in VV124.
Compared to the total luminosity, this gives a value of $M_{HI}/L_V=0.11$, which is particularly low for isolated Local Group dwarfs. The spatial distribution of the gas does not correlate with the observed stellar wings. 
The systemic velocity of the \HI\  in the region superposed to the stellar main body of the galaxy is $V_{\rm h}=-25$~km~s$^{-1}$. The velocity field shows substructures typical of galaxies of this size but no sign of rotation. The \HI\  spectra indicates the presence of a two-phase interstellar medium, again typical of many dwarf galaxies.}


\maketitle
%

\section{Introduction}
\label{intro}

Until just a few years ago, VV124=UGC4879\footnote{A09125+5303 in the nomenclature adopted by \citet{jansen_phot}.} was considered a unassuming isolated dwarf galaxy, classified as spheroidal/irregular,  at a distance of $D\sim 10$~Mpc \citep{halfa}. Integrated multi-color optical photometry was obtained by \citet{jansen_phot} and \citet{taylor}, and J and K$_S$ images were obtained by \citet{nir}, within large surveys of nearby galaxies. From inspection of a low resolution integrated optical spectrum, \citet{jansen_spec} concluded that it was likely ``a young post-starburst galaxy''. 
From the $H_{\alpha}+N[II]$ equivalent width, \citet{jansen_spec} estimated a total star formation rate of $0.005~M_{\sun}$yr$^{-1}$ (assuming distance of 10 Mpc).

The generally adopted distance of $D\sim 10$~Mpc was based entirely on the redshift estimate reported by the CfA survey \citep[$cz=600$~km~s$^{-1}$;][]{cfa}. A team of russian scientists \citep[][K08 hereafter]{k08}, triggered by the apparent partial resolution of VV124 into stars in Sloan Digital Sky Survey images \citep[SDSS,][]{sdss}, carefully searched public databases and the literature, eventually making the case for a much lower recession velocity and a much smaller distance for VV124. They followed up this smart intuition with deep V,I photometry and low-resolution spectroscopy and were actually able (a) to resolve the galaxy into individual stars down to $\sim 2$ mag below the Red Giant Branch (RGB) tip, thus obtaining a direct distance estimate of $D=1.1$~Mpc from the tip itself, and (b) to obtain a new estimate of the heliocentric velocity, much lower than the CfA value, $V_{\rm h}=-70\pm 15$~km~s$^{-1}$ (throughout the paper $V_{\rm h}$ stands for heliocentric radial velocity). This meant that K08  had found a new member of the Local Group (LG), since with the newly determined distances and velocity VV124 is found to lie near the turn-around radius of the LG, and, in fact, being its most isolated member. The galaxy has a remarkable elliptical shape and ranks among the brightest LG dwarf spheroidal/transition type galaxies ($M_B=-11.6$). The stellar budget of the galaxy seems dominated by old stars (RGB; age $\ga 2$~Gyr) with colors compatible with low metallicity (Z$\simeq0.001$). However, a sprinkle of bright blue stars, and the identification of an \HII\ region, led K08 to conclude that VV124 is a transition type between dwarf irregulars (dIrr) and dwarf spheroidals (dSph), like Phoenix, Antlia or LGS3 \citep{mateo}. The results by K08 were further discussed in more detail in
\citet[][T10 hereafter]{tik}.

According to K08, the location and the peculiar velocity of VV124 indicate that it has never been a satellite of a major galaxy of the LG, hence it evolved in full isolation for a Hubble time. Therefore, VV124 may contain a fossil record of precious information on the initial conditions of dwarf galaxies. It may be considered as a
possible progenitor of the gas-less amorphous dSphs found in the vicinity of the Milky Way or M31, whose evolution has been likely largely driven by the strong interaction with the large galaxy they are orbiting around \citep[see][and references therein]{mateo,lokas_struc}. In particular, \citet{lucio1,lucionat} have developed a detailed model, within a modern cosmological context, in which dSphs are produced by the 
morphological transformation of dwarf {\em disk} galaxies by tidal stirring and ram-pressure stripping during their path through the halo of the main galaxy they are gravitationally bound to \citep[see also][and references therein for a more general view on the nature and origin of dSphs]{korme}. As we shall see, VV124 may possibly share some remarkable characteristics with the precursors of modern dSphs envisaged in this model.

The general interest in isolated galaxies as objects of undisturbed evolution is witnessed, for example, by the large Hubble Space Telescope (HST) programme LCID \citep{lcid}, aimed at the determination of the star formation history in the center of six isolated LG dwarfs of various morphological types. Here we are more interested in the structure and dynamics of a galaxy that should be untouched by the interactions with other large galaxies since the beginning of time. In particular, the image presented in Fig.~1 of K08 suggests that the galaxy may be more extended than what could be enclosed into the $6\arcmin \times 6\arcmin$ field studied by those authors. For these reasons, we acquired much deeper observations on a much wider field  with the $2\times 8.4$~m Large Binocular Telescope (LBT, Mt. Graham - AZ). A beautiful color image derived from these data is presented in Fig.~\ref{imaC}, giving also an idea of the number and variety of background galaxies that can be found in such deep LBT images.
In this paper we describe and discuss the results of these observations, as well as those from deep \HI\  data obtained with the Westerbork Synthesis Radio Telescope (WSRT) and from low resolution optical spectroscopy obtained with the Telescopio Nazionale Galileo (TNG). 

The plan of the paper is the following: in Sect.~\ref{phot} we present the LBT observations, we describe the reduction of these data and the artificial stars experiments. The process of surface photometry of the innermost regions of the main body of the galaxy is described, and the adopted system of local coordinates is also introduced. In Sect.~\ref{cmd} we discuss the derived color-magnitude diagrams (CMD), we provide a revised estimate of the distance to VV124 and we analyze the stellar content of the galaxy. Sect.~\ref{struc} is devoted to the analysis of the surface brightness profile and the surface density distribution, while in Sect.~\ref{HIanalysis} the results of the \HI\  observations are discussed in detail; the derived \HI\  velocity field is compared from the velocities obtained from low resolution optical spectroscopy (Sect.\ref{LRS}). Finally, the overall results are summarized and discussed in a broader context in Sect.~\ref{disc}.
 
A few days before this manuscript was ready for submission, a preprint was posted on the {\em astro-ph} archive \citep[][hereafter J10]{jacobs},
presenting deep HST / Advanced Camera for Surveys (ACS) photometry of VV124. This study turns out to be complementary to ours, as it focuses 
on the star formation history (SFH) in the innermost $\simeq 40\arcsec$ of VV124, a region essentially out of reach of our photometry because of the extreme crowding (see Sect.~\ref{comple}). 
We will briefly refer to the results by J10 in the following, when appropriate, but we do not discuss them in detail. In general, for the issues treated in both papers, the results of the two studies are in good agreement.

\section{LBT Observations and data reduction}
\label{phot}

\begin{figure*}
   \centering
   \includegraphics[width=\textwidth]{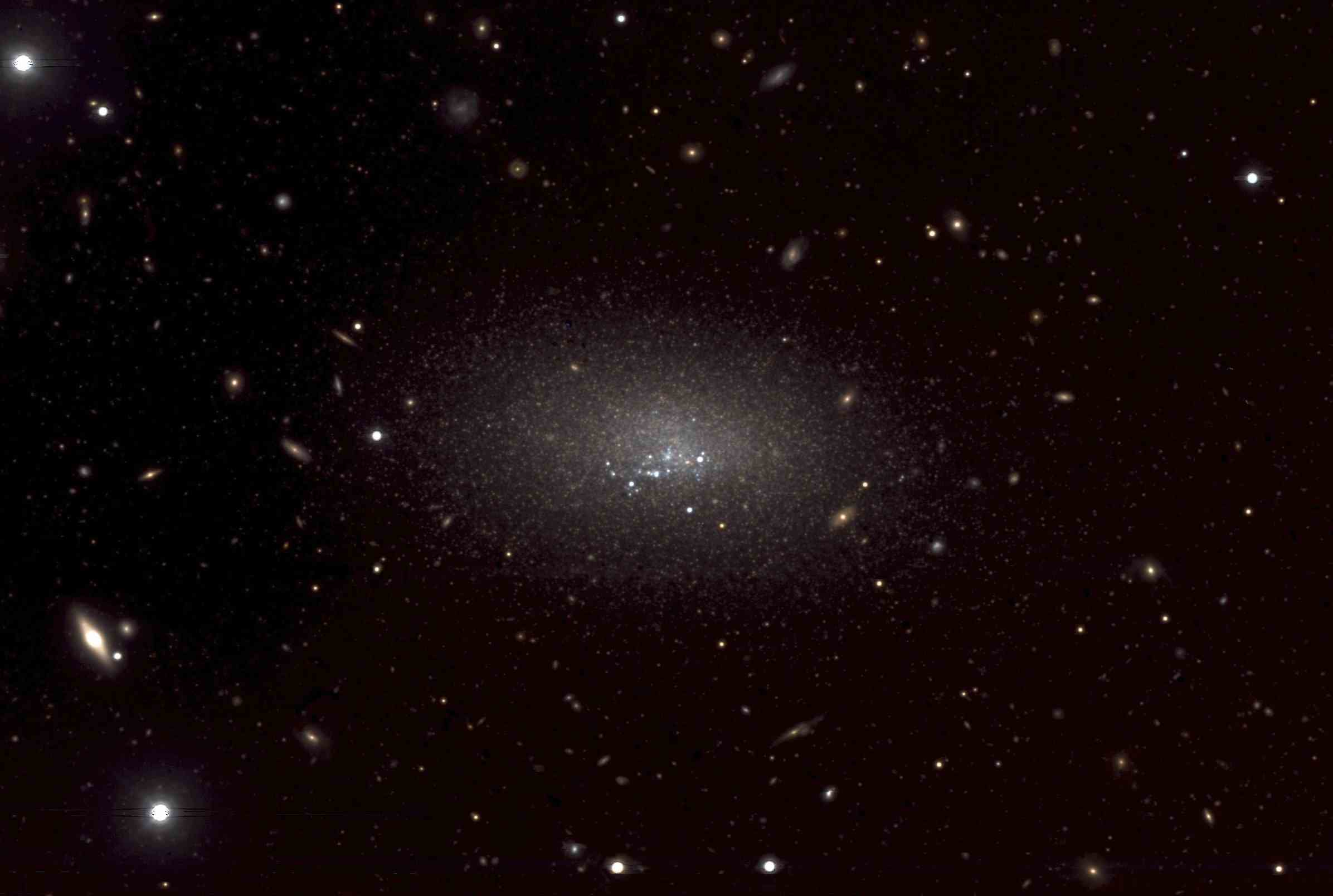}
   \includegraphics[width=\textwidth]{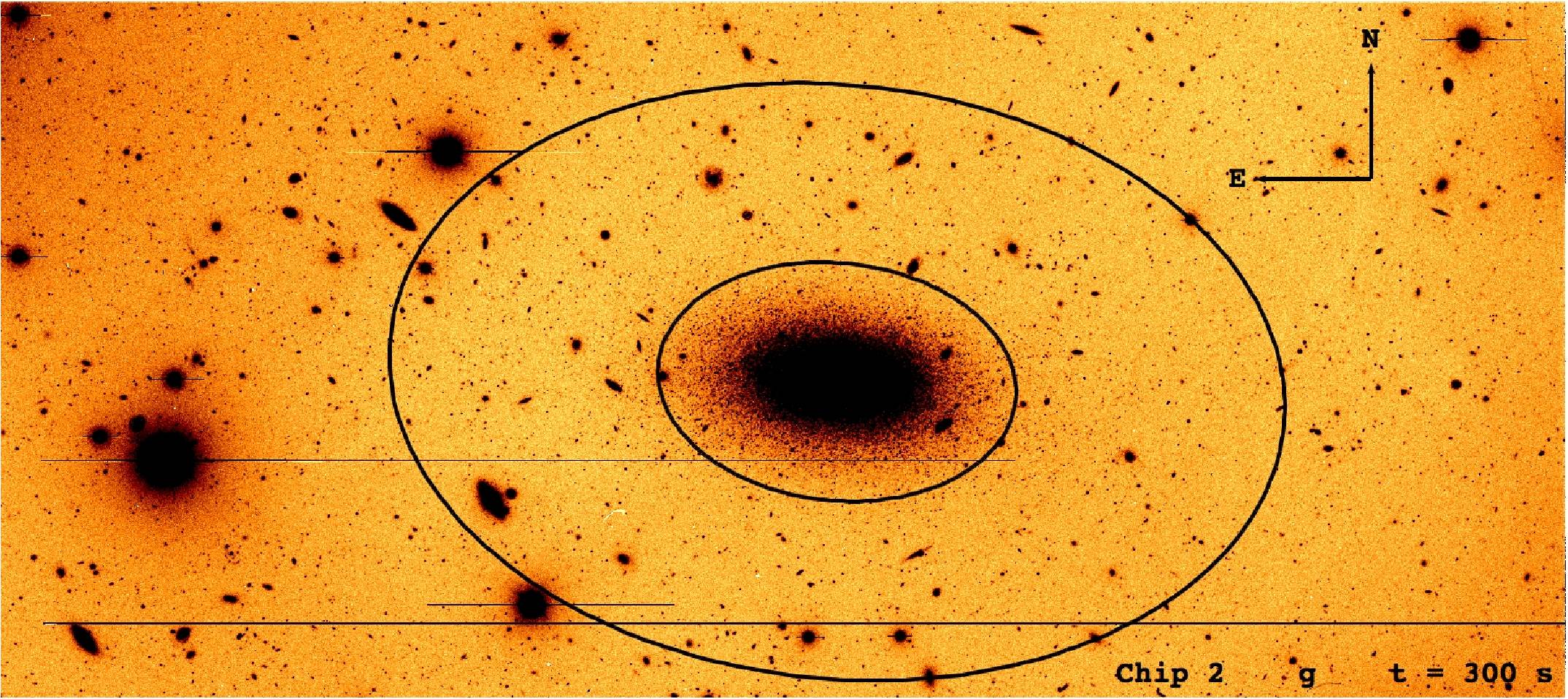}
     \caption{Upper panel: color image of VV124 from the inner $9\arcmin\times 6\arcmin$ of Chip 2. North is up, East to the left.
     Lower panel: $g$ band image of the entire Chip~2, with VV124 near the center. The two   overplotted ellipses have PA=$84.2\degr$ and $\epsilon=0.44$, as derived in Sect.~\ref{sbp}, and semimajor axis of $2\arcmin$ and $5\arcmin$, respectively. }
        \label{imaC}
    \end{figure*}

Deep $g$ and $r$ photometry was acquired on the night of December 2, 2008, at LBT, using the blue channel of the Large Binocular Camera \citep{lbc}.
LBC optics feed a mosaic of four 4608~px~$\times$~2048~px CCDs, with a pixel scale of 0.225 arcsec~px$^-1$. Each CCD chip covers a field of $17.3\arcmin\times7.7\arcmin$. Chips 1, 2, and 3  are flanking one another, being adjacent along their long sides; Chip 4 is placed perpendicular to this array, with its long side adjacent to the short sides of the other chips \citep[see Fig.~4 of][]{lbc}. During our observations the pointing was chosen to place VV124 at the center of Chip~2, with the long side nearly aligned with the major axis of the galaxy (see Fig.~\ref{imaC}). We will consider Chip~1 as our reference control field, sampling the back/foreground population in the direction of VV124; the CMDs from Chip~3 and Chip~4 are undistinguishable from that of Chip~1, hence do not provide any additional information relevant for our purposes. For this reason they will not be discussed anymore in the following.
In the following, we will use the terms Chip~1(2) and field~1(2), abbreviated as f1 and f2, interchangeably.
Several long ($t_{exp} = 300$~s) and short ($t_{exp} = 20$~s) exposures were acquired during the night, but
we selected for the analysis only images taken under excellent seeing conditions, in particular: four $t_{exp} = 300$~s $g$ band images with seeing ranging from $0.60\arcsec$ to $0.69\arcsec$ and two $t_{exp} = 300$~s $r$ band images with seeing ranging from $0.60\arcsec$ to $0.62\arcsec$. Three $g$ and three $r$ $t_{exp}=20$ exposures were also reduced to provide a bridge between the photometry from our long exposures (reaching the saturation level at $r\simeq 19.0$) and the secondary calibrators from the SDSS R6 catalog \citep{sdss} that can be as faint as $g,r\simeq 23.0$ but have average photometric errors $\sigma_g,\sigma_r\le0.03$ mag only for 
$g,r\le 20.0$.

Relative photometry was performed independently on each image using the PSF-fitting code DAOPHOTII/ALLSTAR \citep{daophot,allframe}. Sources with peak higher than 3$\sigma$ above the background were identified in a stacked image obtained by registering and co-adding all the images considered for the analysis. Then, each of these stars was re-identified and fitted on each image (when possible). Only sources found at least in two $g$ {\em and} two $r$ images were retained in the final catalogue. The average and the standard error of the mean of the independent measures obtained from the different images were adopted as the final values of the instrumental magnitude and of the uncertainty on the relative photometry. To clean the catalog from spurious sources, we also adopted cuts in the image quality parameters $CHI$ and $SHARP$, provided by DAOPHOTII. After accurate inspection of the distribution of measured sources in the planes mag vs. $CHI$ and mag vs. $SHARP$, we decided to retain only sources having $CHI \le 2.0$ and $-0.5\le SHARP\le 1.5$. This selection removed   2893 (1185) sources, leaving a final f2(f1) catalog of 15902 (7954) retained sources.

\subsection{Astrometry, Photometric Calibration and Reddening}

The instrumental positions, in pixels, were transformed into J2000 celestial coordinates by means of an astrometric solution (in the form of a first degree polynomial) obtained with CataXcorr\footnote{CataXcorr is a code aimed at cross-correlating catalogues and finding astrometric solutions, developed by P. Montegriffo at INAF - Osservatorio Astronomico di Bologna, and successfully used by our group for the past 10 years.} from 50(34) stars in common between the f2(1) catalog and the SDSS DR6 catalogue; the r.m.s. scatter of the solution  was $0.3\arcsec (0.2\arcsec)$ in both RA and Dec. 

The instrumental magnitudes ($g_i$, $r_i$) were transformed into the SDSS ugriz absolute photometric system ($g_{SDSS}$, $r_{SDSS}$; $g$ and $r$, in the following, for simplicity) with the following equations

\begin{equation}\label{calibg}
g_{SDSS} = g_i +0.103(g_i-r_i) +34.359 
\end{equation}
\begin{equation}\label{calibr}
r_{SDSS} = r_i -0.062(g_i-r_i) +33.911
\end{equation}

\noindent
which were obtained from 43 stars with $g<20.0$ in common between our f2 catalogue and the SDSS DR6 catalog; the r.m.s. is 0.05 mag for both fits, implying an error on the zero point of $\simeq 0.01$ mag. The above equations have been obtained from stars covering the color range $-0.37\le g-r\le 1.55$\footnote{This range includes essentially all VV124 stars observed here (see Fig.~\ref{four}).}.
The calibration of the photometry of the other fields (1, 3, and 4) was obtained by fine adjustments ($\le 0.05$ mag) of the zero point based on the handful of $g<20.0$ stars in common with the SDSS per field, keeping fixed the color coefficients of Eq.~\ref{calibg} and \ref{calibr}. 
During the following analysis, in many instances we will have to compare observables in $g$,$r$  with their counterparts in V,I. To do this we adopt

\begin{equation}\label{traVg}
V=g-0.579(g-r)-0.01~~~~~~~(\sigma=0.005) 
\end{equation}
\begin{equation}\label{traBg}
B =g+0.313(g-r)+0.227~~~~~~~(\sigma = 0.011)
\end{equation}

\noindent
from Lupton (2005\footnote{\tt www.sdss.org/dr4/algorithms/sdssUBVRITransform.html}),

\begin{equation}\label{traIr}
I=r-0.573(g-r)-0.350~~~~~~~(\sigma=0.10)
\end{equation}

\noindent
valid for $g-r<1.4$, and

\begin{equation}\label{tracol}
g-r=-1.417+2.650(V-I)-0.600(V-I)^2~~~(\sigma=0.06),
\end{equation}

\noindent
valid for $0.8\le g-r\le1.4$. Eq.~\ref{traIr} and \ref{tracol} have been derived by us from secondary standard stars in NGC~2419 \citep{n2419}.

We interpolated the \citet{ebv} reddening maps to obtain an estimate of $E(B-V)$ for each source included in our final f1 and f2 catalogues. For both fields we found an average 
$E(B-V)=0.015$ and a standard deviation $\sigma_{E(B-V)}=0.001$. In agreement with previous studies we conclude that in the considered fields the reddening is very low and extremely uniform. In the following we always adopt $E(B-V)=0.015$ and the reddening laws $A_g=3.64E(B-V)$ and $A_r=2.71E(B-V)$, derived by \citet{gira} for cool metal-poor giants.

\subsection{Photometry of unresolved galaxies}
\label{unresolved}

To get some characterization of the background unresolved galaxies which are so abundant in these deep and wide high Galactic latitude fields, we reduced the best $g$ and the best $r$ images with Sextractor \citep{sex}; the astrometric and photometric solutions were transferred from the DAOPHOTII catalog to this new catalog. Sextractor provides total magnitudes, central surface brightness (SB), ellipticity, position angle and other useful parameters for extended objects in astronomical images. Moreover, it provides a "stellarity" index, based on neural network analysis, whose value runs from $s=1.0$ for perfectly point-like sources (stars) to $s=0.0$ for obviously extended sources (galaxies). By inspection of the $g$ vs. $s$ plot we found that the discrimination made by Sextractor between point-like and extended sources is safe down to $g\sim 23.5$ in our images, becoming increasingly blurred at fainter magnitudes.
Hereafter we will refer to the catalog obtained with Sextractor and containing only sources with $s<0.5$ as to the GAL sample, since it should be dominated by background galaxies at any magnitude and it should include almost exclusively galaxies for $g\le 23.5$. 
The GAL sample will be used in Sect.~\ref{cmd} to interpret the background contamination in the CMD of VV124.

\subsection{Artificial stars experiments}
\label{comple}

The completeness of the stellar catalogs has been estimated by means of extensive artificial stars experiments. A total of 100,000 artificial stars have been added to the images \citep[following the recipe described in][]{lf} and the entire data reduction process has been repeated as in the real case, also adopting the same selection criteria described above. The PSF adopted as the best-fit model for photometry was also assumed as the model for the artificial stars. Artificial stars were distributed uniformly in position, over the entire extent of f2, and in color, over the range $-0.6\le g-r\le 1.8$ (see Fig.~\ref{four}, below). They were distributed in magnitude according to a luminosity function similar to the observed 
one but monotonically increasing also beyond the limit of the photometry, down to $r\simeq 27.5$ \citep[see][for details and discussion]{lf}.

   \begin{figure}
   \centering
   \includegraphics[width=\columnwidth]{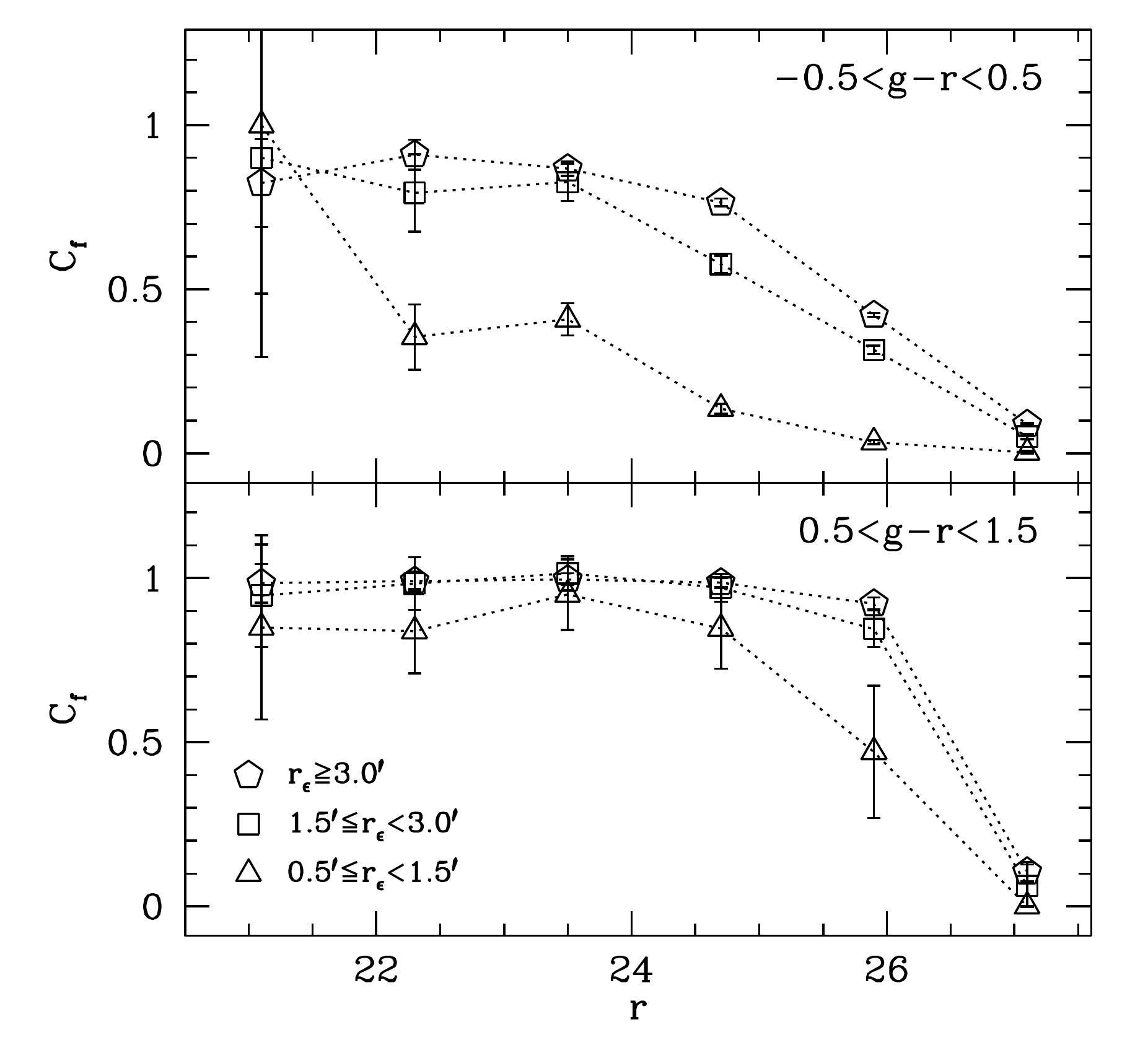}
     \caption{Completeness fraction as a function of r magnitude for different color ranges (upper and lower panels) and for different (elliptical) radial ranges (different symbols).}
        \label{Cf}
    \end{figure}


   \begin{figure}
   \centering
   \includegraphics[width=\columnwidth]{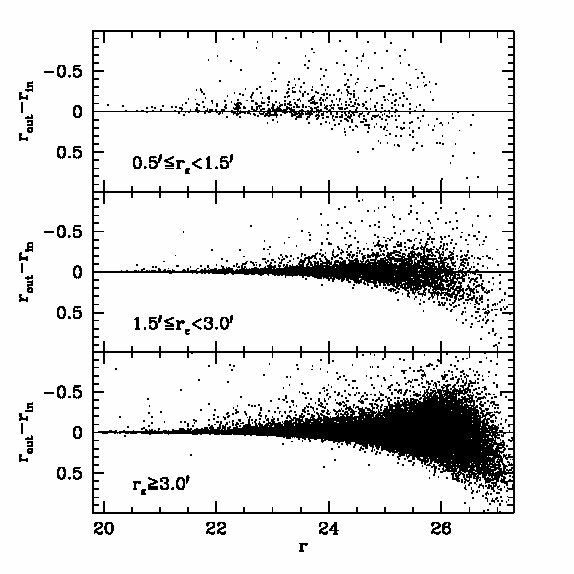}
     \caption{Difference between input and output r magnitudes from artificial stars experiments in three different (concentric) elliptic annuli. The excess of stars with negative difference is due to blends: the recovered star is brighter because it is superposed to another source.}
        \label{blend}
    \end{figure}


In Fig.~\ref{Cf} and Fig.~\ref{blend}, we show the completeness fraction ($C_f$) as a function of r magnitude for different color ranges and for different (elliptical) radial ranges (see Sect.~\ref{sbp}, for a definition of $r_{\epsilon}$), and the
difference between input and output r magnitudes  for different (elliptical) radial ranges, respectively. It is interesting to note that in the color range covered by the RGB of VV124 ($0.5\le g-r<1.5$, see Fig.~\ref{four}), the completeness does not change with $r_{\epsilon}$ for $r_{\epsilon}> 1.5\arcmin$; in this range $C_f$ is fairly constant and larger than 80\% for $r\le 26.0$.
Finally, we note also that for  $r_{\epsilon}<0.5\arcmin$ the completeness is lower than 50\% virtually at any magnitude and the photometric errors due to crowding/blending becomes very significant. In agreement with these findings, in the observed catalogue there are only 129 stars in this radial range, nearly all brighter than $r\simeq 24.5$. For these reasons, in the following we will limit our analysis to stars having 
$r_{\epsilon}\ge 0.5\arcmin$.

\subsection{Surface photometry and coordinate system}
\label{sbp}

\begin{figure}
   \centering
   \includegraphics[width=\columnwidth]{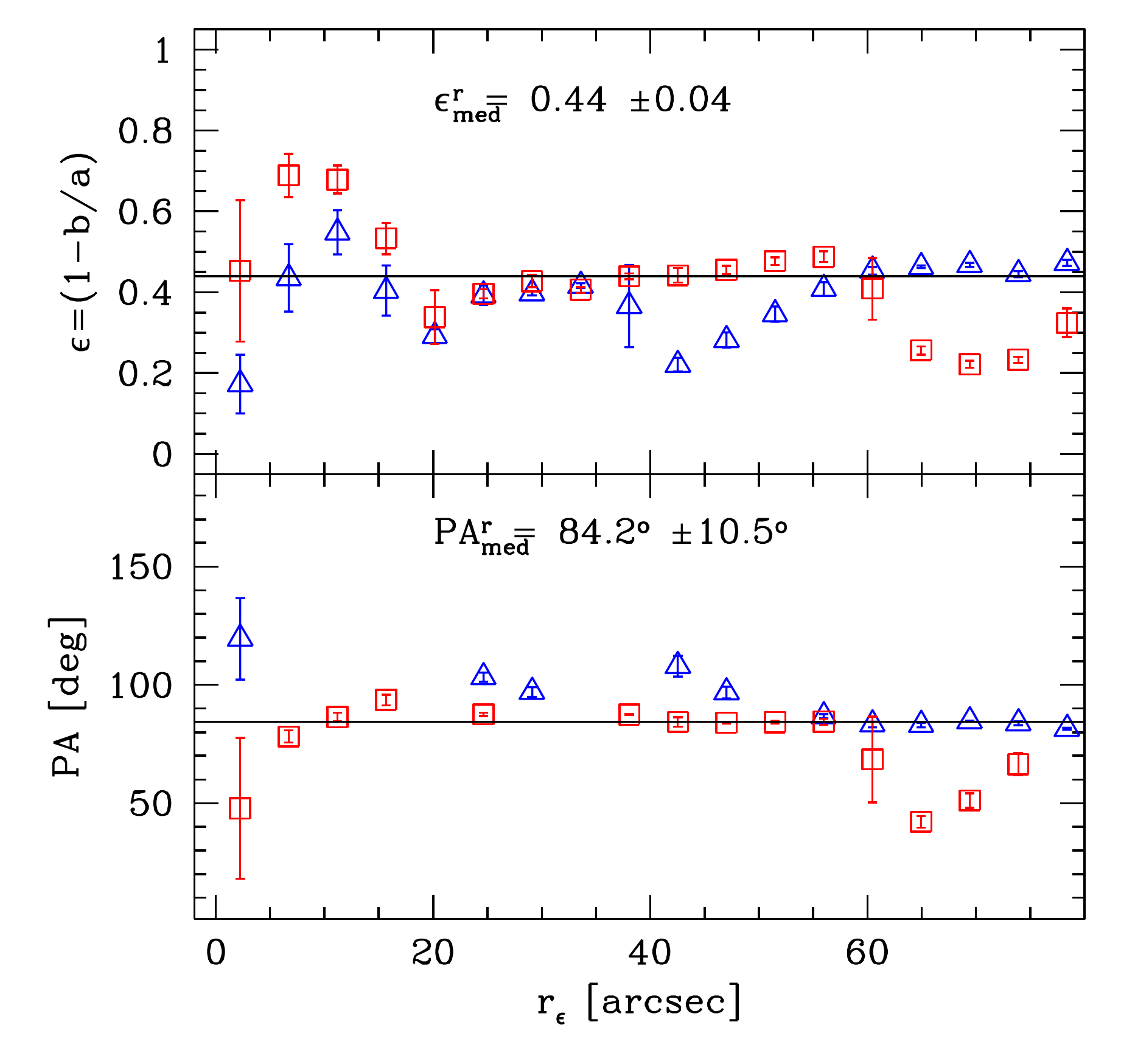}
     \caption{Ellipticity and position angle as a function of distance from the galaxy center along the major axis, from XVISTA surface photometry. Red open squares are from the analysis of the r image, blue open triangles are from the $g$ image. Each plotted point is the mean of the considered quantities over a radial range of $\pm 10$~px$\simeq \pm 2.2\arcsec$ around the position of the point, the errorbars are the associated standard deviations \citep[as in][]{lucky}. In the lower panel we plotted only points having uncertainties in PA lower than $30\degr$. Continuous horizontal lines mark the average levels of the parameters considered in each panel; the corresponding values are also reported.}
        \label{ellpa}
    \end{figure}

We performed surface photometry on the innermost $72\arcsec$ of the best $g$ and $r$ images using
XVISTA\footnote{\tt http://astronomy.nmsu.edu/holtz/xvista/index.html} 
\citep[see][for details on the code and on the adopted procedure]{xvista,lucky}.
The derived surface brightness profile will be discussed in Sect.~\ref{popstar}, below. 
For the purposes of the present section, we show in Fig.~\ref{ellpa} the radial profiles of the ellipticity ($\epsilon$) and of the position angle obtained with XVISTA by fitting ellipses to the observed light distribution (PA, measured anti-clockwise from north towards  east). Over the considered radial range, both quantities are remarkably constant with radius, independent of adopted passband. For this reason, we adopted the average values ($\pm$ the associated standard deviation) as our final estimates of $\epsilon$ and PA, in particular $\epsilon=0.44\pm 0.04$ and $PA=84.2\degr \pm 10.5\degr$, in good agreement with the values reported by K08, \citet{jansen_spec} and \citet{taylor}.

To obtain the coordinates of the center of the smooth elliptical light distribution of the main body of the galaxy (i.e. not affected by the asymmetric distribution of the young stars), 
we obtained a smoothed intensity contour within the central   
$\simeq 1\arcmin$ from the $r$ and $g$ images; next we fitted these contours with ellipses having $\epsilon=0.44$ and $PA=84.2\deg$, and we took the average of the coordinates of the centers of the ellipses fitted to the $g$ and $r$ images as our final center coordinates ($\alpha_0$,$\delta_0$)\footnote{The final estimates are the result of the iteration of the described process: first XVISTA has been run by adopting a center position estimated by eye, thus obtaining an estimate of 
$\epsilon$ and $PA$, then the center was estimated by fitting ellipses and XVISTA was run again with the newly determined center position. The process converged after the first iteration, i.e. the $\epsilon$ and $PA$ profiles were essentially unchanged by the adoption of new coordinates of the center.}. 
The $g$ and $r$ centers differ by $ 1.3\arcsec$; we took the average of the two as our final result. Our center is $11.2\arcsec$ different (approximately in the E-N-E direction) from the coordinates provided by NED (and reported by K08 in their Tab.~1) and $\sim 6\arcsec$ west from the center plotted by K08 in their Fig.~2. Tests on several images convinced us that our $\alpha_0$,$\delta_0$ provide the best approximation for the center of the overall light distribution, as shown in the lower panel of Fig.~\ref{imaC}, above.
Adopting these coordinates for the center, we
convert to cartesian coordinates $X$,$Y$(in arcmin) projecting the equatorial
coordinates of each star ($\alpha$,$\delta$) on the plane of sky as 
in \citet{ven} 

\begin{displaymath}
X\prime=-(10800/\pi)\cos(\delta)\sin(\alpha-\alpha_0)
\end{displaymath}
\begin{displaymath}
Y\prime=(10800/\pi)[\sin(\delta)\cos(\delta_0)-\cos(\delta)\sin(\delta_0)\cos(\alpha-\alpha_0)]
\end{displaymath}

\noindent
with $X\prime$ increasing toward west and $Y\prime$ increasing toward north. These 
$X\prime,Y\prime$ coordinated were then rotated by $90\degr-PA$ into a $X$,$Y$ system where the $X$ axis is parallel to the major axis (and $Y$ parallel to the minor axis) of the galaxy.
Finally, we defined the elliptical distance from the center of the galaxy (or elliptical radius $r_{\epsilon}$) as:

\begin{displaymath}
r_{\epsilon}=\sqrt{X^2+{\Big(\frac{Y}{1-\epsilon}\Big)}^2}
\end{displaymath}

\noindent
that is equivalent to the major-axis radius. The $X$,$Y$ coordinate system and $r_{\epsilon}$ will be always adopted in the following analysis.

\begin{table}
  \begin{center}
  \caption{Observed and derived parameters of VV124}\label{Tab_par}
  \begin{tabular}{lcr}
    \hline
    Parameter & value & notes\\
\hline
$\alpha_0$     & 09:16:03.18                    & J2000  \\
$\delta_0$     & +52:50:31.3                    & J2000 \\
$l_0$          & $164.6619\degr$                  & Gal. long.\\
$b_0$          & $42.8864\degr$                   & Gal. lat.\\
SGL            & $47.6121\degr$                 & Supergal. long.\\
SGB            & $-15.0110\degr$                & Supergal. lat.\\
E(B-V)         & $0.015\pm 0.001$               &   \\
$(m-M)_0$      & $25.61\pm 0.13$                &   \\
D              & $1.3\pm 0.1$ Mpc               &   \\
1~arcsec       & 6.3~pc                         & conv. factor at D=1.3~Mpc\\
$\epsilon$     & $0.44\pm0.04$                  &   \\
PA             & $84.2\degr \pm 10.5\degr$      &   \\
$\mu_r(0)$     & $21.1 \pm 0.10$ mag/arcsec$^2$ & $^a$  \\ 
$\mu_g(0)$     & $21.0 \pm 0.10$ mag/arcsec$^2$ & $^a$   \\ 
$\mu_V(0)$     & $21.2 \pm 0.15$ mag/arcsec$^2$ & $^a$   \\ 
$R_S^r$        & $17.4\arcsec$                  & S\'ersic scale radius  \\
$R_S^V$        & $17.5\arcsec$                  & S\'ersic scale radius  \\
$R_e^r$        & $41.3\arcsec$                  &  $^b$                       \\
$\mu_r(R_{\epsilon}<R_e)$      & $22.5 \pm 0.15$ mag/arcsec$^2$ &  $^{a,c}$ \\ 
$r_h$          & $39.7\arcsec$                  &  $^d$ \\
$g_{tot}$      & $13.5 \pm 0.10$                &   \\
$r_{tot}$      & $13.0 \pm 0.10$                &   \\
$B_{tot}$      & $13.8 \pm 0.15$                &   \\
$V_{tot}$      & $13.2 \pm 0.15$                &   \\
$M_V$          & $-12.4 \pm 0.20$               &  \\
$L_V$          & $8.2^{+1.6}_{-1.4}\times 10^6~L_{V,\sun}$    & total V luminosity \\
$V_{\rm h}$          & $-25\pm 5$~km~s$^{-1}$                 & $^e$ \\
$V_{g}$        & $+17$~km~s$^{-1}$              & galactocentric velocity$^f$ \\
$M_{stars}$    & $1.6\pm 0.3\times 10^7~M_{\sun}$      & stellar mass$^g$  \\
$M_{HI}$       & $8.7\times 10^5~M_{\sun}$      & gaseous mass  \\
\hline
\end{tabular}
\tablefoot{$a$ {\em Observed} values, not corrected for extinction (note that the correction is smaller than the uncertainty on SB in all the 
passbands considered here).
$^b$ $R_e^r$ is the elliptical radius containing half of the galaxy light.$^c$ $\mu_r(R_{\epsilon}<R_e)$ is the SB within the ellipse of that radius; note that this is the quantity that must be used in Fundamental Plane plots, not the SB {\em at} $R_e$, see \citet{gadotti}.
$^d$ $r_h$ is the radius of the {\em circular} area containing half of the galaxy light; the same value is consistently obtained from both $g$ and $r$ images.
$^e$ Mean systemic heliocentric radial velocity derived from \HI\  observations.
$^f$ Adopting the solar motion from \citet{schon} and $V_{rot}=220$~km~s$^{-1}$.
$^g$ Adopting ${\frac{M}{L_V}}=2.0$.
B an V magnitudes have been obtained from their $g$ and $r$ counterparts using equations
\ref{traBg} and \ref{traVg}.} 
\end{center}
\end{table}

\section{Stellar content and Distance}
\label{cmd}
\subsection{The Color Magnitude Diagram}

\begin{figure*}
   \centering
   \includegraphics[width=\textwidth]{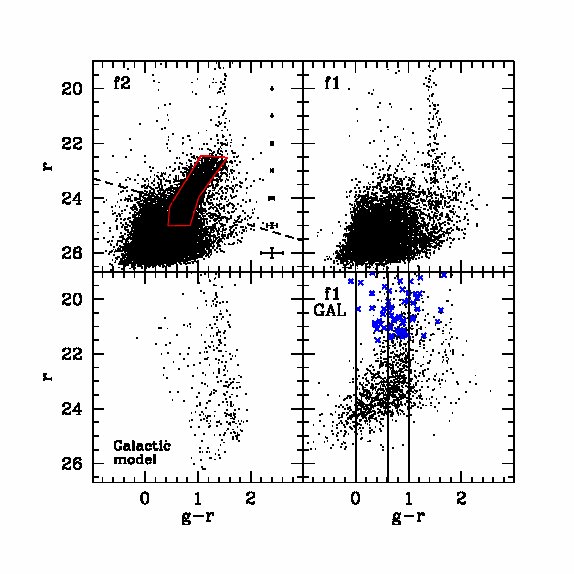}
     \caption{Upper panels: CMDs of the entire Chip~2 (left) and Chip~1 (right) fields. The dashed line marks the limiting magnitude reached in the V,I photometry by K08. The average $r$, $g-r$ uncertainties are plotted as errorbars. 
     The comparison between the two panels reveals the steep and wide RGB of VV124, tipping at $r\simeq 22.5$, that dominates the f2 CMD while is completely lacking in the f1 one.
     The red polygon enclosing the RGB of VV124 is used to select stars for the star counts described in Sect.~\ref{struc}.
     Lower left panel: predictions of the TRILEGAL Galactic model \citep{trilegal} for a field of the same area as f1 and f2 in the direction of VV124. The model includes {\em only} Galactic stars. The effects of photometric uncertainties have been properly added. Lower right panel: 
     CMD of the sources classified as {\em galaxies} by Sextractor in f1 (small dots). The blue $\times$ symbols are the sources brighter than r=20.5 classified as {\em galaxies} in the SDSS catalogue in the same area. The vertical lines approximately enclose the color range spanned by {\em Red} and {\em Blue} Sequence galaxies, according to \citet{blanton}; see also \citet{balogh}.}
        \label{four}
    \end{figure*}

The total CMDs of f2 and f1 are compared in the upper panels of Fig.~\ref{four}. Our new photometry reaches $r=26.5$, a full 2 mag deeper than that of K08.
The RGB of VV124 is the dominant feature in the f2 CMD, running from 
(g-r, r)$\simeq (0.5,26.5)$ to (g-r, r)$\simeq (1.2,22.5)$ and having no counterpart in f1.
The comparison reveals that the prominent vertical plume lying to the blue of the RGB, around $(g-r)=0$,
is equally present in f1 and f2, indicating that it is not due to a population that is characteristic of VV124. 

In this context, it is worthwhile interpreting the features that can be ascribed to
fore/background contaminating populations which can be identified in the right panel of Fig.~\ref{four}. To obtain an idea of the foreground contamination by Galactic stars, we run a simulation with the TRILEGAL Galactic model \citep{trilegal}. The synthetic TRILEGAL CMD for a FoV of the same area as f1 and f2 is shown in the lower left panel of Fig.~\ref{four}; the effects of photometric errors and completeness have been properly included. The model reproduces remarkably well the nearly vertical narrow plume observed in any field at $g-r\sim 1.2$. This is the well-known red plume of local M dwarf stars. The sparse diagonal band of stars going from 
(g-r,r)$\sim(1.1,26.0)$ to (g-r,r)$\sim(0.2,19.0)$ is made of Main Sequence (MS) stars lying at different distances in the Galactic halo. Also this sequence is clearly identified in the CMDs of f1 and f2, for $r\la 23.0$, where it is not hidden by other populations. The feature may appear as remarkably narrow and well defined in the f2 CMD, but a simple inspection of the spatial distribution of the involved stars reveals they are clearly not correlated with VV124, confirming their Galactic origin.


The above discussion unavoidably leads to the conclusion that for $r\ga 23.5$ and $g-r\la 1.2$ the contaminating population is completely dominated by distant galaxies. This is a well known characteristic of deep photometric surveys \citep[see, for example][their Fig.~6, in particular]{ibam31}. This interpretation is strongly supported by the CMD of the GAL sample, shown in the lower right panel of Fig.~\ref{four}: all the f1 sources plotted here have been classified as galaxies by Sextractor (small dots) or by the SDSS pipeline ($\times$ symbols). They form two broad sequences in color matching very well the ranges spanned by the well-known red and blue sequences of galaxies \citep[see][for references and discussion]{blanton}.

\begin{figure}
   \centering
   \includegraphics[width=\columnwidth]{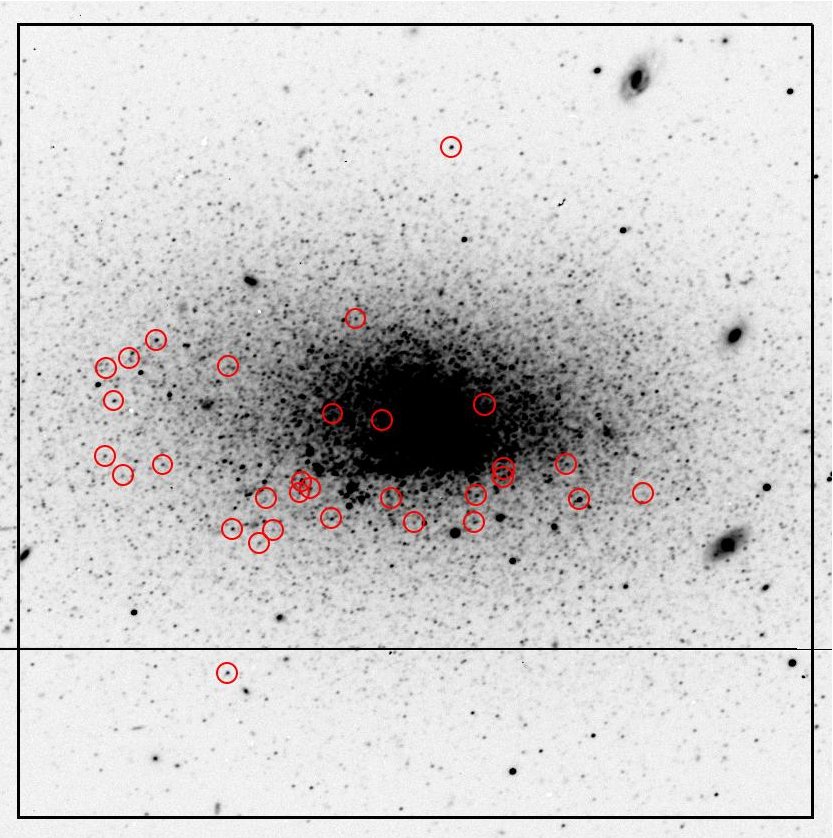}
     \caption{Sources having r$<25.0$ and $g-r\le -0.3$ are overplotted (red open circles) on a LBC $g$ image of VV124 zoomed on the center of the galaxy. The main body of the galaxy is enclosed within a square of side $l=3\arcmin$. The orientation is the same as in Fig.~\ref{imaC}.}
        \label{BPima}
    \end{figure}

A minor population of young Main Sequence / Blue Loop stars is very likely hidden in the strong background contamination, in this region of the CMD, since we see the bluest stars of this population emerging from the blob of contaminating galaxies at $g-r\le -0.3$ (see also J10). K08 identified a small population of bright MS stars as young as a few tens of Myr, displaying an asymmetric distribution with respect to the smoothly elliptical shape of the overall galaxy (see Fig.~\ref{imaC}). In particular, they found that the brightest blue stars and one \HII\ region are confined into a relatively narrow sheet in the southern half of the central $30\arcsec$ of the galaxy. Most of the $g-r\le -0.3$ stars detected here in f2 are fainter than $I\sim 24$, thus they were out of reach of the K08 photometry. Figure~\ref{BPima} suggests that they are likely the faint counterpart of the bright blue stars discussed by K08: it is clear that they are associated with VV124 and the majority of them are confined in to the same strip located $\sim 10-20\arcsec$ to the South of the galaxy center.  We used the synthetic CMD technique \citep[see][for references]{cignoni} to reproduce the observed number 
of these young stars (limiting the analysis to $r\le 25.0$), with the main aim of obtaining a rough estimate of the star formation rate (SFR) at recent epochs. The total mass of stars with age $\le 500$ Myr is 
$\sim 1.4\times10^5~M_{\sun}$, assuming a Salpeter Initial Mass Function and a metallicity Z=0.001, i.e. $\sim \frac{1}{100}$ of the total stellar mass of the galaxy (see Tab.~\ref{Tab_par}). The resulting average SFR $\sim 0.0003~M_{\sun}$~yr$^{-1}$ is in reasonable agreement with that obtained by \citet{halfa} from the integrated $H_{\alpha}$ flux, once rescaled to the correct distance of VV124 (1.3~Mpc instead of 10.5~Mpc), i.e. SFR=$0.00008~M_{\sun}$~yr$^{-1}$, given the considerable uncertainties involved. The agreement is good also with the SFH derived by J10 in the same age bin.

While the presence of these stars and their asymmetric distribution indicate  recent activity in the galaxy (possibly connected to the asymmetric structure of the \HI\  density and velocity field, see Sect.\ref{HIanalysis}), the associated star formation episode produced only a minor component of the overall stellar mix populating VV124. This young population is more abundant in the innermost region of the galaxy and is best characterized with the HST photometry by J10, who concluded that it is quite rare at distances larger than $40\arcsec$ from the center and that it formed in the last 500~Myr. The same is true for the {\em very} sparse population of candidate AGB stars brighter than the tip. 
For these reasons, in the following we will focus mainly on the RGB stars that are the visible part of the old stellar population which seems to dominate the galaxy.

\subsection{Distance from the tip of the RGB}
\label{distance}

   \begin{figure}
   \centering
   \includegraphics[width=\columnwidth]{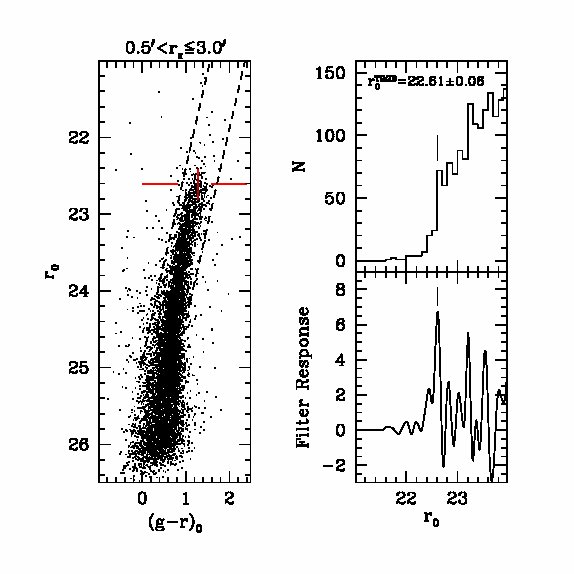}
     \caption{Determination of the r magnitude of the RGB tip. Left panel: de-reddened CMD for the indicated elliptical annulus. The two parallel dashed lines enclose the RGB stars selected to derive the luminosity function. The pair of horizontal segments mark the level of the tip, the vertical segment marks the average color at the tip. Upper right panel: LF of the RGB. The sharp cut-off corresponding to the TRGB is marked by a vertical segment. Lower right panel: Sobel filter response as a function of r magnitude. The peak corresponding to the TRGB is marked by a vertical segment.}
        \label{tip}
    \end{figure}


The tip of the red giant branch \citep[TRGB;][]{lfm93} is widely recognized as one of the most
straightforward and general applicable distance indicators \citep[see][B08 hereafter, for references and discussion]{cefatip}. In the present case, it is also the only standard candle accessible with our data to estimate the distance of VV124. Figure~\ref{tip} shows the very clean detection of the TRGB of VV124 we obtained from our data: the tip is unequivocally detected by the Sobel filter \citep[see][B08]{lfm93} as an obvious cut-off in the luminosity function (LF) of the RGB at $r^{TRGB}_0=22.61\pm 0.06$, where the reported uncertainty is the half width at half maximum (HWHM) of the highest peak in the filter response (lower right panel of Fig.~\ref{tip}). It is interesting to note that the uncertainty of the TRGB position is a factor of 2 lower than reported in T10, owing to the much higher resolution/accuracy of our own photometry. The average color of the RGB at the tip is $(g-r)_0=1.28 \pm 0.02$.

While an empirical calibration of the absolute magnitude of the TRGB in the SDSS passbands is still lacking, in B08 and \citet{tipleiden} it has been shown that all available theoretical models agree in predicting a very modest dependency of $M^{TRGB}_{r}$, $M^{TRGB}_{i}$, and  $M^{TRGB}_{z}$ on metallicity and age, at least for relatively metal-poor models ($[Fe/H]\le -0.7$; this is also the main reason which makes $M^{TRGB}_{I}$ so appealing as a standard candle, B08). In particular, considering \citet{gira} models of age=~12~Gyr and 4~Gyr and spanning the full range of available metallicity, we found that for TRGB colors $(g-r)_0\le 1.4$ the average magnitude of the tip is $M^{TRGB}_{r}=-3.03$ with a standard deviation of just 0.04 mag. This is in excellent agreement with the predictions of the  independent set of isochrones by \citet[][see B08 for some example in other passbands]{dotter}. The low scatter in $M^{TRGB}_{r}$ in the considered range of age and colors is confirmed also by BASTI \citep{basti} models, but in this case the average magnitude of the tip is $\simeq 0.08$ mag brighter than in other models \citep[see][]{tipleiden}. The latter difference is still within the uncertainty in the absolute zero point of the calibration \citep[see][B08 and references therein]{tip2}. Adopting $M^{TRGB}_{r}=-3.03\pm 0.04$, $r^{TRGB}_0=22.61\pm 0.06$ and including an uncertainty of 
0.1 mag for the overall zero point we obtain $(m-M)_0=25.61 \pm 0.13$; taking BASTI models we would have obtained $(m-M)_0=25.72$, instead.

Another route to estimate the distance from the tip is to convert the observed magnitude and colors from $r_0, (g-r)_0$ to $I_0, (V-I)_0$ using the relations reported in Sect.~\ref{phot} and then to adopt the empirical calibration:

\begin{equation}
\label{tipcal}
M^{TRGB}_I = 0.080(V-I)_0^2-0.194(V-I)_0 -3.939  ~~~\pm 0.12
\end{equation}

derived by B08 from the original calibration as a function of [Fe/H] obtained in \citet{tip1} and revised in \citet{tip2}. We obtain $(V-I)_0^{TRGB}=0.981\pm 0.06$ and, in turn,
$M_I^{TRGB}=-4.05 \pm 0.12$, that coupled with $I_0^{TRGB}=21.53\pm 0.10$ gives $(m-M)_0=25.58\pm 0.16$, fully supporting the conclusions reached above, based on the \citet{gira} and \citet{dotter} models. As our final best estimate we adopt $(m-M)_0=25.61 \pm 0.13$, corresponding  to $D=1.3\pm 0.1$~Mpc; this value will be always adopted in the following. At this distance 1 arcmin corresponds to 378 pc. 

The main reason for the marginal difference with respect to the estimate by T10 ($D=1.1\pm 0.1$~Mpc) resides in the brighter level of the TRGB found by T10: this may be due (a) to a larger impact of blending in their lower resolution photometry, and/or (b) to a systematic error in the absolute photometric calibration which should not affect our photometry as we have the standards in field. Further support for the larger distance estimate obtained here is provided by the excellent agreement with the estimate by J10\footnote{J10 reports $(m-M)_0=25.67 \pm 0.04$. Note that the reported error does not account for the systematic uncertainty in the zero point of the calibration, that is of order 0.1 mag \citep[][B08]{tip2}.}.

\subsection{Metallicity}
\label{popstar}

\begin{figure}
   \centering
   \includegraphics[width=\columnwidth]{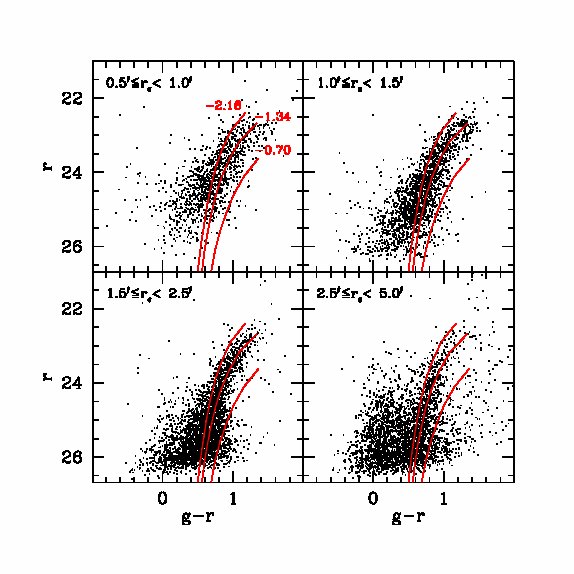}
     \caption{RGB fiducial ridge lines are super-imposed to the CMD of stars in four different elliptical
     annuli around the center of  VV124.
     The ridge lines are labeled with the metallicity of the associated cluster (in the CG97 scale) in the upper left panel.}
        \label{rad}
    \end{figure}

In Fig.~\ref{rad} we compare the CMD of VV124 in different elliptical annuli with a grid of RGB ridge lines of Galactic globular clusters from the set by \citet{clem}, converted from the
$u\prime g\prime r\prime i\prime z\prime $ system to the $ugriz$ one according to \citet{tuck}.
From blue to red, the RGB templates are for: M92, at [Fe/H]=-2.16 \citep[all the metallicities are in the scale by][CG97 hereafter]{cg97}, M3 ([Fe/H]=-1.34) and M71 ([Fe/H]=-0.70).
The reddening and distance moduli adopted for the templates are taken from \citet{f99}. 
The comparison of the observed RGB with GC templates is the standard way to obtain estimates of the metallicity of an old stellar population from photometric data \citep[see, e.g.][and references therein]{harrimet,leo2}. It is known to provide valuable constraints on the overall metallicity distribution when the considered RGB is dominated by stars with age comparable with that of classical GCs ($\ga 8-10$~Gyr\footnote{See \\{\tt www.lorentcenter.nl/lc/2009/324/friday/Bellazzini.ppt} for some recent tests on the reliability of the technique.}). The SFH derived by J10 from HST data indicates that the vast majority of VV124 stars are older than 10~Gyr even in the innermost region, thus supporting the idea that the analysis is well suited for the present case .

The first general remarks that can be made based in inspecting Fig.~\ref{rad} are : (a) the majority of RGB stars of VV124 are remarkably metal-poor, with $[Fe/H]\le -1.0$, and (b) the observed color spread at any magnitude is significantly larger than expected from mere photometric errors, hence there should be a significant spread in metallicity (see below); both conclusions were drawn also by K08 and are confirmed by J10. The distributions of RGB stars
in the four panels seem to indicate that the innermost regions host a larger fraction of relatively metal-rich stars compared to the external ones, in agreement with the typical metallicity gradient observed in dSphs \citep{harbeck}. 
It may be hypothesized that the larger number of RGB redder than the M3 ridge line in the innermost annulus is due to a larger fraction of blending in this most crowded region (see Fig.~\ref{blend}). However, this cannot be the case, as blending spuriously increase the luminosity of stars, thus mimicking bluer colors and {\em lower} metallicities, not higher. On the other hand, higher crowding also implies larger photometric errors and, consequently, broader color distributions in the inner regions. 
Tests performed using synthetic stars from our set of artificial stars experiments suggests that it is very unlikely that the observed changes of the color distribution of RGB stars with radius is due to mere observational effects. 

We obtained a rough estimate of the metallicity of each star by linear interpolation on the grid of ridge lines of Fig.~\ref{rad}, as done, for example, in \citet{harrimet}, \citet{savi}, and \citet{micmet}. To limit any kind of contamination as well as the impact of photometric uncertainties we considered only the stars having $22.7\le r<24.0$ and enclosed between the ridge lines of M92 and M71. The average metallicity and the standard deviation in the outermost radial bin considered in Fig.~\ref{rad}, i.e. the one less affected by crowding and photometric errors, are $\langle [Fe/H\rangle=-1.5$ (which we adopt as the typical metallicity of the galaxy, in the following) and $\sigma_{[Fe/H]}^{obs}=0.34$ dex. Once de-convolved from the dispersion due to observational effects ($\sigma_{[Fe/H]}^{synth}=0.13$ dex, estimated from artificial stars), this leads to an intrinsic one-sigma metallicity spread of $\sim 0.3$ dex, typical of dSph galaxies \citep{mateo,tht}.

\subsubsection{An extended Horizontal Branch?}
\label{hbsect}

\begin{figure}
   \centering
   \includegraphics[width=\columnwidth]{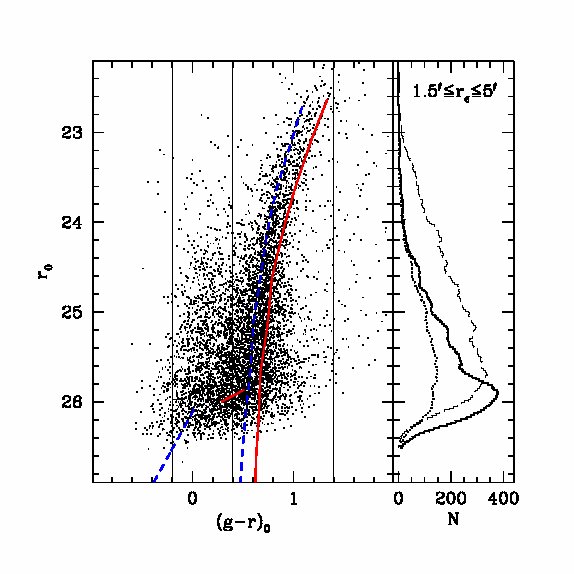}
     \caption{Left panel: de-reddened CMD of the stars lying in the
     elliptical annulus $1.5\arcmin\le r_{\epsilon}\le 5.0\arcmin$, with the (RGB and HB) ridge lines of the GCs NGC2419 (dashed blue line) and Pal4 (continuous red line), de-reddened and shifted to the distance of VV124, as derived here. 
     The ridge lines has been derived by us from public SDSS data.
     The thin vertical lines enclose the typical color range of HB ($-0.2<(g-r)_0\le 0.4$) and RGB ($-0.4<(g-r)_0\le 1.4$) stars; the latter range is expected to enclose also the Red Clump of an intermediate-age population. Right panel: the LF of the stars in the HB color range in the considered annulus (thick continuous line) is compared to the LF in the same color range from an equivalent area of f1 (dotted line). The thin long-dashed line is the LF of the stars in the RGB color range.}
        \label{hb}
    \end{figure}

J10 clearly detect an RC (and an old Red HB) approximately peaking at a magnitude corresponding to our $r_0 \sim 26.5$; they attribute part of this population to stars with age between 0.5 and 1 Gyr. However, this is only a minor component of their SFH, hence the majority of the core-He-burning stars they observe must be associated with the dominant old population. We do not see any sign of an RC  population, even in the less-crowded outer regions of VV124 ($r_{\epsilon}\ge 1.5\arcmin$). The LF of the stars
in the color range expected to enclose such a feature is shown as a long-dashed thin line in the right panel of Fig.~\ref{hb} and it shows no hint of the characteristic peak due to an RC population \citep[compare, for example, with the case of Leo~I, which hosts a significant intermediate-age population,][]{leo1,sdssleo}. This is likely due to the sudden drop of completeness occurring in our sample in this magnitude range, although radial age/metallicty gradients may also play a role. Probably the RC is less conspicuous in the outer regions of VV124, if they are populated by older and more metal-poor stars than those inhabiting the center, on average. 

Along these lines, it is suggestive to note, in Fig.~\ref{hb}, that the HB ridge lines of the Galactic globular clusters NGC2419 ($[Fe/H]=-2.1$, \citet{harris}; distance and reddening from \citet{ripepi}) and Pal4 ($[Fe/H]=-1.4$, \citet{p4}; distance and reddening from \citet{harris}), nicely fall on top an over-density that has no counterpart in the CMD of the control field (f1). 
This is clearly shown in the right panel of 
Fig.~\ref{hb} where the LF of VV124 stars in the color range $-0.2<(g-r)_0\le 0.4$ is compared with stars in the same range from a field of equal area from f1. While the LF of the control field rise gently up to $r\sim 26$ where it drops down to the limiting magnitude, the LF of the considered annulus around VV124 shows a sharp rise of star counts at $r\simeq 25.6$, peaking at $r\simeq 25.9$\footnote{All the results presented in this section are unchanged if completeness-corrected LFs are considered instead of the uncorrected ones shown in Fig.~\ref{hb}.}.

We tentatively interpret this feature as an extended old HB population associated with VV124. If this interpretation is correct, the galaxy must host a sizable population of RR Lyrae: unfortunately, the entire set of observations used in this paper was acquired in less than one hour, making it impossible to search for candidate variable stars.
It must be recalled that all the results presented in this sub-section concern stars within one magnitude from the limit of our photometry, i.e. a realm where the completeness is rapidly dropping and the photometric errors are quite large.
Yet we feel that they are worth to be reported, in the hope they may trigger further investigations.

\section{Structure}
\label{struc}

\begin{figure}
\centering
\includegraphics[width=\columnwidth]{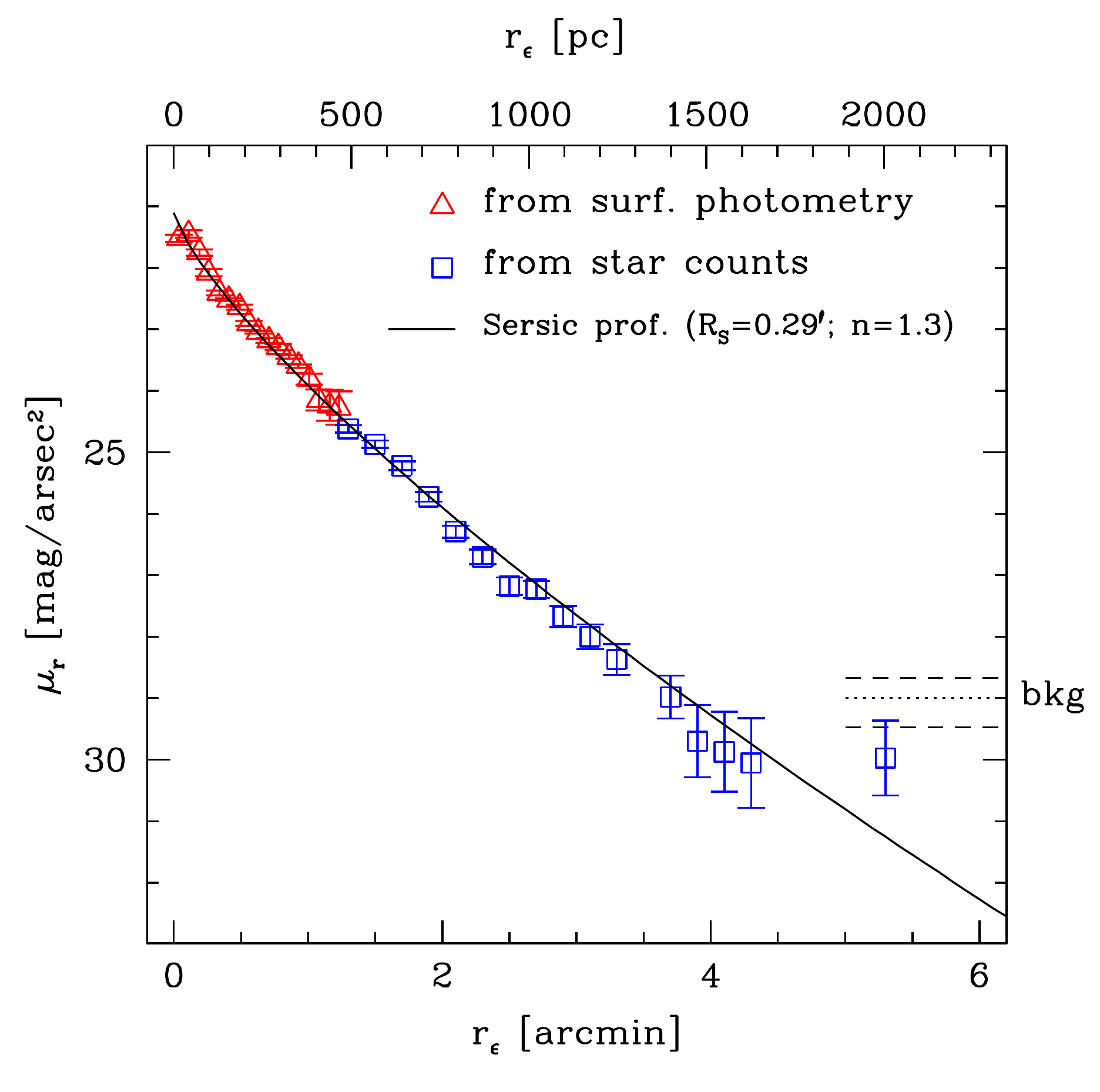}
\caption{Surface brightness profile of VV124 in r band obtained by joining surface photometry (for $r_{\epsilon}\le 72\arcsec$; Sect.~\ref{sbp}) and star counts (for $r_{\epsilon}> 72\arcsec$) profiles. The continuous line is the Sersic model that best-fits the surface photometry. The level of the background (bkg, dotted line) and the associated uncertainty (dashed lines) are also reported. The error bars from surface photometry are typically smaller than the dimension of the points.}
        \label{prof}
    \end{figure}

In Fig.~\ref{prof} we show the azimuthally-averaged major-axis r-band surface brightness (SB) profile of VV124. The profile has been obtained by joining the surface photometry on concentric elliptical apertures obtained with XVISTA out to $r_{\epsilon}=72\arcsec\simeq 1.2\arcmin$ (see Sect.~\ref{sbp}), with the surface density profile obtained from star counts  in $r_{\epsilon}$\footnote{I.e., keeping fixed the values of $\epsilon$ and $PA$ derived in Sect.~\ref{sbp}.}, which are equivalent to star counts on elliptical annuli \citep[see][for a discussion of the procedure and details]{lucky}. For star counts we used likely VV124 members selecting candidate RGB stars enclosed in the polygonal box shown in the upper left panel of Fig.~\ref{four}. Star counts are limited to the largest ellipse (with the parameters reported in Sect.~\ref{sbp}) which is completely enclosed within f2, i.e. only to RGB stars having $r_{\epsilon}\le 6.0\arcmin$.
The two profiles overlap with two points in the region between $\simeq 1\arcmin$ and $\simeq 1.2\arcmin$: the overlap region was used to normalize the star counts profile, shifting it to the same scale of SB of the surface photometry profile\footnote{For $r_{\epsilon}\le 1\arcmin$ the star counts profile is strongly affected by radial variations in the completeness factor. The surface photometry profile was transformed onto the standard scale of $g,r$ magnitudes using Eq.~\ref{calibg} and \ref{calibr}.}. The SB values (not corrected for extinction) as a function of $r_{\epsilon}$ are listed in Tab.~\ref{sbprof}.

Surface brightness profiles of VV124 were previously obtained (in several passbands) by \citet{jansen_phot} and by \citet{taylor}, from surface photometry limited to $r_{\epsilon}\le 1.5\arcmin$. Once transformed to the proper photometric band (using Eq.~\ref{traVg} and \ref{traBg}), both profiles are in good agreement with ours in the overlapping region.
The novelty of Fig.~\ref{prof} is that using star counts in the outer regions we were able, for the first time, to trace the SB profile out to $\simeq 5\arcmin \simeq 1.9$~kpc, reaching $\mu_r\simeq 30.0$~mag/arcsec$^2$, thus demonstrating that the galaxy is much more extended than previously believed. It is interesting to note the a \citet{sersic} model of scale-length $R_S=0.29\arcmin = 110$~pc and $n=1.3$ (i.e., a nearly exponential profile, in agreement with T10) provides a good fit to the observed profile over the whole considered range $0.0\arcmin \le r_{\epsilon}\le 5.0\arcmin$. 

The main structural parameters we obtained from the observed profile are listed in Tab.~\ref{Tab_par}. To estimate the integrated magnitude, we computed the total flux within the maximum elliptical aperture we reach with surface photometry 
($r_{\epsilon}=72\arcsec$); next we used the best-fit S\'ersic model to find the fraction of light outside that radius. It turned out that a correction of $-0.1$~mag should be applied to the magnitude within $r_{\epsilon}=72\arcsec$ to account for the contribution of these external regions. By transforming our integrated $g$ and $r$ magnitudes into an integrated B magnitude (with Eq.~\ref{traBg}) we obtain $B_{tot}=13.8\pm 0.25$ in good agreement with previous estimates from \citet[][reported also by K08]{taylor}  and \citet{jansen_phot}, $B_{tot}\simeq 13.7$.
The absolute V magnitude is $M_V=-12.4$, corresponding to $L_V=8.2\times10^6~L_{V,\sun}$.

\begin{table}
  \begin{center}
  \caption{Observed Surface Brightness profiles of VV124.}
  \label{sbprof}
  \begin{tabular}{cccc}
$r_{\epsilon}$ & $\mu_r$ & $\mu_g$ & source\\
arcmin         & mag/arcsec$^2$ & mag/arcsec$^2$ &  \\
\hline
  0.04 & $21.52 \pm   0.06$  & $21.40 \pm   0.02$  & sp\\ 
  0.11 & $21.45 \pm   0.06$  & $21.75 \pm   0.07$  & sp\\ 
  0.19 & $21.75 \pm   0.05$  & $21.94 \pm   0.05$  & sp\\ 
  0.26 & $22.08 \pm   0.06$  & $22.30 \pm   0.06$  & sp\\ 
  0.34 & $22.41 \pm   0.04$  & $22.59 \pm   0.06$  & sp\\ 
  0.41 & $22.52 \pm   0.02$  & $22.67 \pm   0.06$  & sp\\ 
  0.49 & $22.64 \pm   0.04$  & $22.96 \pm   0.05$  & sp\\ 
  0.56 & $22.90 \pm   0.04$  & $23.18 \pm   0.04$  & sp\\ 
  0.63 & $23.05 \pm   0.02$  & $23.48 \pm   0.04$  & sp\\ 
  0.71 & $23.18 \pm   0.04$  & $23.94 \pm   0.05$  & sp\\ 
  0.78 & $23.29 \pm   0.05$  & $23.99 \pm   0.04$  & sp\\ 
  0.86 & $23.45 \pm   0.03$  & $24.06 \pm   0.03$  & sp\\ 
  0.93 & $23.60 \pm   0.03$  & $24.12 \pm   0.03$  & sp\\ 
  1.01 & $23.81 \pm   0.09$  & $24.17 \pm   0.03$  & sp\\ 
  1.08 & $24.15 \pm   0.17$  & $24.34 \pm   0.02$  & sp\\ 
  1.16 & $24.24 \pm   0.25$  & $24.54 \pm   0.02$  & sp\\ 
  1.23 & $24.28 \pm   0.27$  & $24.79 \pm   0.02$  & sp\\ 
  1.30 & $24.62 \pm   0.06$  & $24.92 \pm   0.06$  & sc\\ 
  1.50 & $24.87 \pm   0.06$  & $25.17 \pm   0.06$  & sc\\ 
  1.70 & $25.22 \pm   0.07$  & $25.52 \pm   0.07$  & sc\\ 
  1.90 & $25.72 \pm   0.08$  & $26.02 \pm   0.08$  & sc\\ 
  2.10 & $26.29 \pm   0.10$  & $26.59 \pm   0.10$  & sc\\ 
  2.30 & $26.70 \pm   0.12$  & $27.00 \pm   0.12$  & sc\\ 
  2.50 & $27.18 \pm   0.14$  & $27.48 \pm   0.14$  & sc\\ 
  2.70 & $27.23 \pm   0.14$  & $27.53 \pm   0.14$  & sc\\ 
  2.90 & $27.67 \pm   0.17$  & $27.97 \pm   0.17$  & sc\\ 
  3.10 & $28.00 \pm   0.20$  & $28.30 \pm   0.20$  & sc\\ 
  3.30 & $28.37 \pm   0.25$  & $28.67 \pm   0.25$  & sc\\ 
  3.70 & $28.98 \pm   0.35$  & $29.28 \pm   0.35$  & sc\\ 
  3.90 & $29.70 \pm   0.59$  & $30.00 \pm   0.59$  & sc\\ 
  4.10 & $29.87 \pm   0.65$  & $30.17 \pm   0.65$  & sc\\ 
  4.30 & $30.05 \pm   0.73$  & $30.35 \pm   0.73$  & sc\\ 
  5.30 & $29.97 \pm   0.61$  & $30.27 \pm   0.61$  & sc\\ 
\hline
\end{tabular} 
\tablefoot{sp = surface photometry; sc = star counts.}
\end{center}
\end{table}

\subsection{Density maps}
\label{dmap}

\begin{figure}
   \centering
   \includegraphics[width=\columnwidth]{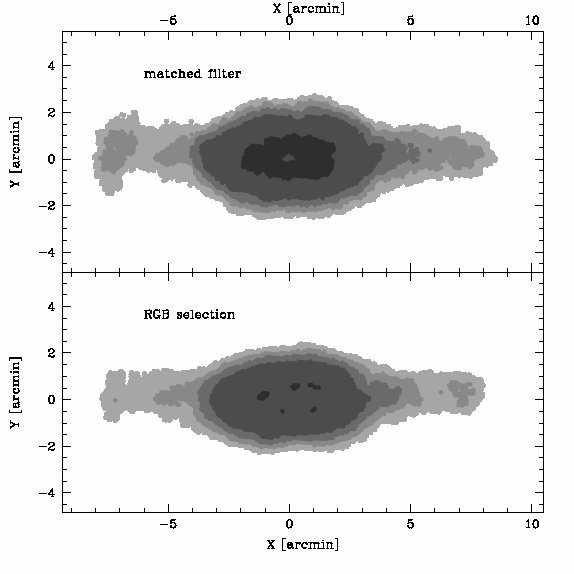}
     \caption{Upper panel: density map obtained with the {\em matched filter} (MF) technique, adopting the pattern of weights shown in greyscale in Fig.~\ref{pesi}. Lower panel: density map obtained using RGB stars selected to lie within the box shown in Fig.~\ref{pesi}. The levels of grey range between $3\sigma$ to $8\sigma$ above the background, in steps of $1\sigma$, from the lightest to the darkest tone of grey. The density depressions near the center of the main body are due to the low degree of completeness in the very crowded innermost 
     $r_{\epsilon}\le 0.5\arcmin$ region.}
        \label{maps}
    \end{figure}

\begin{figure}
   \centering
   \includegraphics[width=\columnwidth]{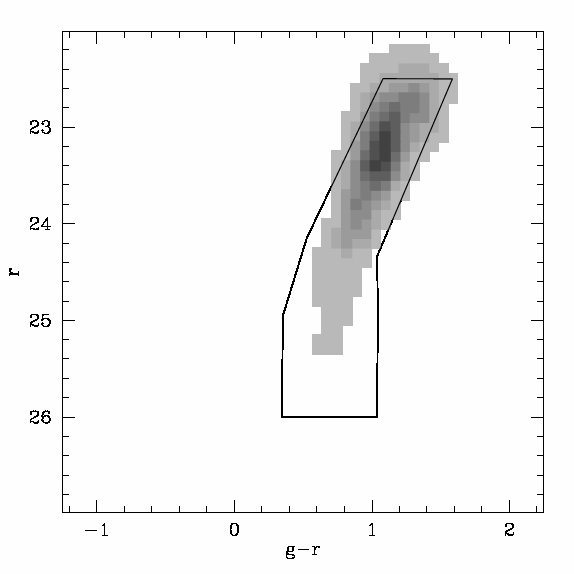}
     \caption{Selection box (continuous line) and weight pattern (in greyscale) adopted for the density maps shown in Fig.~\ref{maps}. Higher weights corresponds to darker grey areas, going from 0.1 (lightest grey) to 1.0 (darkest grey), with step 0.1. Weight = 0.0 is assigned to stars outside the grey area, hence these stars do not contribute to the density estimate.}
        \label{pesi}
    \end{figure}

The two dimensional density maps shown in Fig.~\ref{maps} are even more interesting than the extended profile shown above. Both maps are obtained by estimating the density from star counts on a fixed regular grid of nodes spaced by $0.2\arcmin$ in both directions, with an adaptive algorithm that adjusts the spatial resolution according to the local density. 
To describe the algorithm, it is more convenient to
refer to the map in the lower panel of Fig.~\ref{maps}. In this case, likely RGB 
stars of VV124 are selected by requiring they lie within the selection box plotted in Fig.~\ref{pesi}, similarly to what was done for the SB profile. Next, for each node of the grid we find the distance $D_{50}$ of the 50th star in order of distance from the node itself. Following this,  the local density (in stars/arcmin$^2$) is computed by dividing 50 by the area of the circle of radius $D_{50}$ \citep[see][for discussion and another application of the same concept]{silv,umi}. The average density of the background, computed using the same selection over the whole f1, is then subtracted to this number and the total 1-$\sigma$ error of the 
background-subtracted density is computed, assuming Poisson statistics. In the maps of Fig.~\ref{maps} the density is expressed in units of $\sigma$ above the background.

The map in the upper panel has been obtained in the same way, but adopting the matched filter (MF) scheme \citep{connie} instead of a selection box. At each grid node the density is computed by summing the {\em weights} of the 50 stars within $D_{50}$; the weights are assigned according to the star position in the CMD within the weight pattern shown in greyscale in Fig.~\ref{pesi}. The weights are obtained as a map of ratios of the number of stars from f2 to the number of stars from f1 in a given position on the CMD, normalized to have the maximum equal to unity. It is clear from Fig.~\ref{pesi} that the MF approach is more conservative, as the bulk of the signal is provided by the most likely genuine VV124 RGB members, most of the weight being assigned to stars brighter than $r=24.5$. This is the reason the associated density map is less smooth than the one derived from the selection box, which equally weights all the candidate RGB stars down to $r=26.0$. In any case, we want to stress that we have obtained density maps using various adaptive and non-adaptive (fixed radius) density estimators and {\em in all cases we obtained density distributions with the same features shown in Fig.~\ref{maps}}.

The most striking features of the density maps are two symmetric, relatively thin wings emanating from the edges of the elliptical (bulge-like\footnote{Or {\em pseudo-bulge}-like \citep[see, e.g.,][and references therein]{gadotti}.}) main body around $X\ga 3\arcmin$ and reaching $X\simeq 8\arcmin$, approximately aligned along the major axis of the galaxy. These structures have no obvious counterpart in the other dwarf galaxies of the Local Group. The overall density distribution recalls a disk galaxy with a prominent bulge, seen nearly edge-on. 
The galaxy is quite gas-poor as a whole and no hint of \HI\  is observed in the wings (see 
Sect.~\ref{HIanalysis}). Hence,  the wings would be interpreted as the dry remnant of an ancient disk that evolved undisturbed until the total consumption of the original gas it was made of. In this context, it is interesting to recall that T10 interpret the structure of VV124 as a superposition of disks of different thickness, the thickest being populated by old RGB stars, the thinnest by young Main Sequence and Red Super Giants.
Alternatively, the wings may be identified with the inner regions of two tidal tails \citep[see][and references therein]{munoz,klim_tid}, but this explanation seems unlikely, given the extreme isolation of VV124 from any other mass distribution (Both the MW and M31 are $\simeq 1.3$~Mpc away, and the nearest dwarf is Leo~A, at a distance of $\simeq 700$~kpc; see J10). 

   \begin{figure}
   \centering
   \includegraphics[width=\columnwidth]{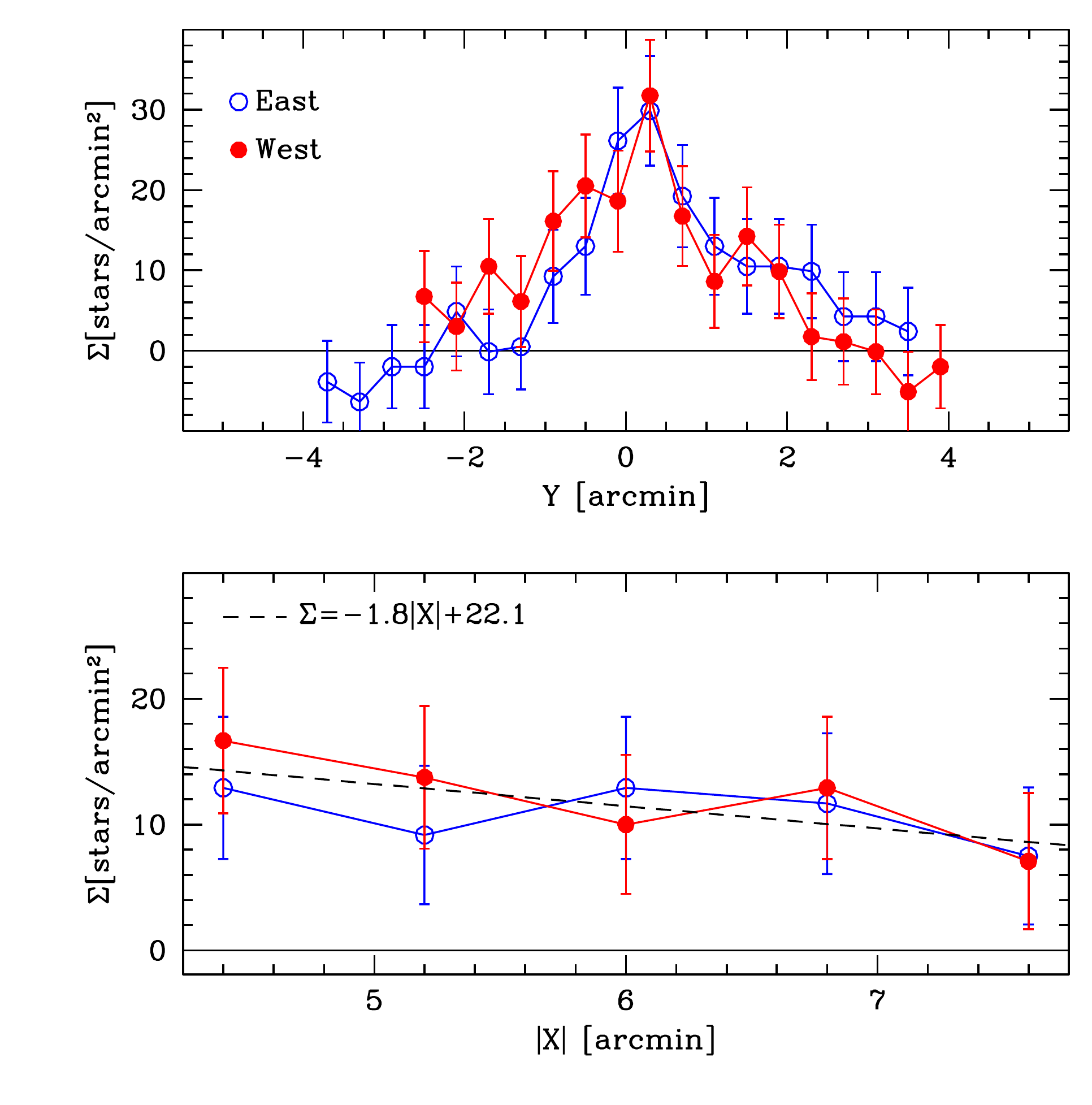}
     \caption{Background-subtracted surface density profiles of the Eastern (blue open circles) and Western (red filled circles) wings of the density distribution shown in Fig.~\ref{maps}, from counts of RGB stars selected with the same box used for the density maps (see Fig.~\ref{pesi}). Upper panel: profiles along the Y direction. Lower panel: profiles along the 
     $|X|$ direction; the adoption of the absolute value of $X$ allows to compare the Eastern to the Western wing in the same scale. The dashed line is the best linear fit to the distribution of points from both wings.
     In all the cases only stars within $|Y|\le1.5\arcmin$ and $4\arcmin\le |X|\le 8\arcmin$ have been considered.}
        \label{wings}
    \end{figure}


To further check the significance of these very unusual structures, as well as their actual association to VV124, we obtained the surface density profiles of the two wings along the $X$ and $Y$ directions, using stars selected to lie within the RGB box shown in Fig.~\ref{pesi}. In both cases, the background has been estimated from f1, adopting the same selections and procedures, and subtracted to the f2 profiles.
Since the major axis of the galaxy is inclined by just $5.8\degr$ with respect to the E-W direction, in the following we will refer to the wing at positive $X$ as to the Western wing and to that at negative $X$ as to the Eastern wing. In Fig.~\ref{wings} the $Y$ (upper panel) and $|X|$ (lower panel) profiles of the Eastern and Western wings are compared. There are several conclusions that can be drawn from these plots:

\begin{itemize}

\item The over-density in the considered regions is clear and significant in both profiles and in both wings. 

\item The profiles of the E and W wings are indistinguishable, both in the $Y$ and in the $|X|$ direction. Both $Y$ profiles have a clear peak near $Y=0$, i.e. at main the plane of symmetry of the whole galaxy.

\item 
We do not see any obvious truncation of the wings in the $X$ direction, but it is hard to draw any conclusion on this, given the very low surface brightness of the considered structures. Extrapolating the linear fit to the observed profiles (dashed line), $\Sigma=0$ is reached at $|X|\sim 12\arcmin$. Note that the fit is not intended to bear any physical meaning, it was attempted just to perform the above extrapolation to zero surface density.

\end{itemize} 

From the results described above it can be concluded that the wings are clearly associated to VV124, as it is very hard to conceive how any kind of unrelated structure in the background can present such a degree of symmetry in the Western and Eastern sides of the galaxy, as well as such a degree of correlation with the major axis of the galaxy. 

From this analysis it is not possible to draw firm conclusions on the actual nature of the wings. However, it should be noted that the strong peaks at $Y=0$ shown by the density profiles in the $Y$ direction seems more typical of a disk than of a tidal stream.
The kinematics of stars in the whole galaxy {\em and} in the wings will certainly provide very useful insight into the origin of the wings, e.g., by looking for the signature of coherent rotation about the minor axis \citep[see, for example, the case of the isolated Tucana and Cetus dSph][respectively]{tuc,cetus}.

To obtain further insight into the possible tidal  origin of the wings, we performed simple plausibility tests by comparing the observed size of VV124 to its expected
tidal radius for two basic scenarios, described below.
The tidal radius $r_t$ is defined as the cut-off radius in the density distribution of a system of mass $m$ imposed by the tides from a mass $M$ lying at distance $D$. In this case we will use the formula

\begin{equation}
\label{RT}
r_t=\frac{2}{3}\left(\frac{m}{2M}\right)^{\frac{1}{3}} D
\end{equation}

which is appropriate for a logarithmic potential 
\citep[see][for discussion and references]{tidal}. 

\begin{enumerate}

\item Let us consider the tidal force exerted on VV124 by the Local Group as a whole, as if the LG was a single giant galaxy whose center is located at the barycenter of the LG, adopted to lie at $D=1.2$~Mpc from VV124 \citep[][]{vdb00}. This is clearly too simple a model for the mass distribution of the LG, but it may provide a rough upper limit to the tidal force that may be felt by VV124. Taking the stellar mass listed in Tab.~\ref{Tab_par} as the total mass of VV124 (a conservative choice, in this context) we find that for the range spanned by the most recent estimates of the LG mass \citep[$2-6\times10^{12}~M_{\sun}$, see][and references therein]{vandermarel} the expected tidal radius ranges between 8~kpc and 13~kpc. This is significantly larger than the maximum size of the main body of VV124 ($r_{\epsilon}\simeq 2$~kpc), and also larger than the size of the wings themselves, which reach $r_{\epsilon}\simeq 3$~kpc. Hence, it seems unlikely that the structure of VV124 could have been shaped by the tidal interaction with the LG as a whole or by one of its major components, as both the MW and M31 are more distant and less massive than the ideal super-galaxy considered here. 

\item We now test the case of a close interaction with a dwarf galaxy. In the hierarchical paradigm, the accretion of small groups of dwarf galaxies by a major halo should not be uncommon 
\citep[see][and references therein]{donghia}, thus such a scenario is not unlikely, in principle. We consider the nearest dwarf to VV124, i.e. Leo~A, a dIrr galaxy with mass $M\simeq 8\times10^7~M_{\sun}$ \citep{brown}. Solving Eq.~\ref{RT} for D, we found that for Leo~A to impose a tidal radius as small as 3~kpc to VV124, the two dwarfs should have been as close as $\sim 10$~kpc in the past. It is likely that from such a close encounter the two galaxies should have emerged either as a bound pair or with high peculiar velocities \citep{sales}, none of which is currently observed. 

\end{enumerate}

To follow up on these simple considerations, we plan to search for isolated dwarfs showing signs of tidal interactions in recent simulations of the evolution of LG-size groups, performed in a realistic cosmological context. However, this is clearly beyond the scope of the present paper.

\section{Neutral Hydrogen}
\label{HIanalysis}

In order to study the possible presence of neutral hydrogen in VV124, the galaxy was observed with the WSRT on April 13, 2008.   A bandwidth of 10 MHz (corresponding to $\sim 2000$ km s$^{-1}$) was used which was covered with 2048 channels, resulting in a velocity resolution of about 2 km s$^{-1}$ after Hanning smoothing the spectra.  The integration time was 12 hr. Calibration and analysis of the data were done following standard WSRT recipes using \textit{MIRIAD} software \citep{Sault95}.  Several datacubes were made by varying weighting and tapering. Lower resolution cubes, which, in principle, are more sensitive to faint, extended emission, do not show more \HI\  compared to the full resolution datacube. The spatial resolution of this datacube is $13\times16$ arcseconds and the noise level is 1.2 mJy beam$^{-1}$. The faintest detected emission  is at the level of $5 \times 10^{19}$ cm$^{-2}$.

\begin{figure}
\centering
\includegraphics[width=\columnwidth]{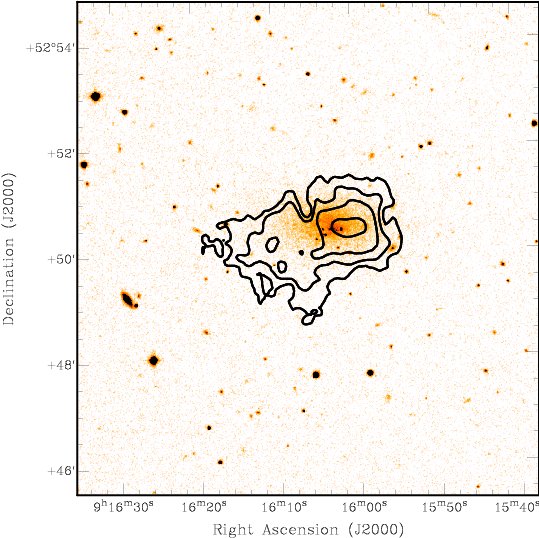}
\caption{Total \HI\  contours on top of an optical image (with image cuts putting in evidence only the densest central part of the stellar body of the galaxy). Contour levels are 5, 10, 20 and $50\times10^{19}$ cm$^{-2}$.}
\label{totHI}
\end{figure}

Neutral hydrogen was clearly detected in VV124.   Figure \ref{totHI} shows the distribution of  \HI\  in relation to the optical galaxy. The total \HI\  image was derived by smoothing the full-resolution data to 30 arcsec resolution and clipping this smoothed datacube at 3 $\sigma$ to make a mask which was applied to the full resolution datacube. The \HI\  is fairly well centered on VV124, with the peak of the emission about 1 arcmin W of the center. To the SE the \HI\  extends to larger radius and seems to form a tail-like structure. 

The flux integral of the \HI\  emission is $2.2\pm 0.1$ Jy km  s$^{-1}$ which, for a distance of 1.3 Mpc, gives a total \HI\  mass of $8.7\pm 0.4 \times~10^5\ M_\odot$. Given the optical luminosity of VV124, this implies that, although this galaxy contains a significant amount of \HI, it is relatively gas poor. In a study of Local Group dwarf galaxies, \citet{Grcevich09} find that Local Group dwarfs that are, like VV124, about 1 Mpc distant from the Milky Way and Andromeda typically have  $M_{\rm HI}/L_V \simeq 1$, while for VV124  $M_{\rm HI}/L_V = 0.11$. Only the Cetus and Tucana dwarfs are more gas poor (with $M_{\rm HI}/L_V$ below 0.01).

\begin{figure}
\centering
\includegraphics[width=\columnwidth]{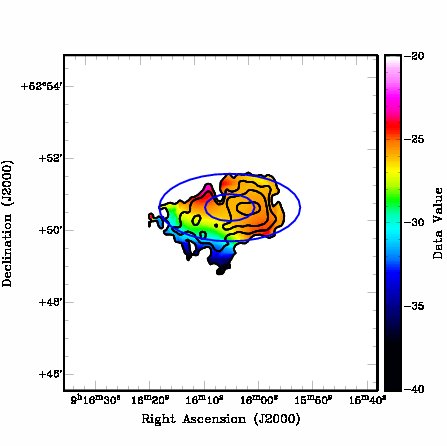}
\caption{Velocity field of the \HI\  emission of VV124 (colourscale, heliocentric velocity in units of km s$^{-1}$) with overlaid the contours of the total \HI\  (same contour levels as in Fig.~\ref{totHI}). The superposed blue ellipses are centered on the optical center, has $\epsilon=0.44$ and have the semi-major axis equal to the optical effective radius (inner ellipse) and a=$2\arcmin$, corresponding to the $\mu_r=26.0$~mag/arcsec$^2$ isophote.}
\label{velfie}
\end{figure}

In Fig.~\ref{velfie} we show the velocity field of the \HI\  emission. This velocity field was derived using the intensity weighted mean of the masked cube that was also used for deriving the total \HI\  emission. The kinematics of the \HI\  does not show any signature of overall rotation over the optical body of VV124. Such kinematics is seen in most dwarfs with a dynamical mass below $10^8\ M_\odot$ \citep{Begum08}. The systemic velocity of the \HI\  coinciding with the optical body is $V_{\rm h} = -25\pm 4$ km s$^{-1}$, quite different from $V_{\rm h} \simeq -70$ km s$^{-1}$ reported by K08; a detailed comparison between the \HI\  velocity field and the results from optical spectroscopy is reported in Sect.~ \ref{LRS}.
The tail-like extension to the SE seems to have a velocity gradient towards more negative velocities. 

 The origin of this \HI\ tail is not clear. A possibility would be that the gas is displaced from the main body by the combined action of stellar winds. 
The small gradient in velocity ($5-10~{\rm km\,s^{-1}}$) and the small mass of gas {\em in the tail} ($\simeq 10^5~M_{\sun}$) would require a kinetic energy of $\sim 10^{51}$ erg, at most. Such an energy can be provided by about 10 OB stars within a few million years with an efficiency of transferring kinetic energy to the ISM of 10\%.
However, if the \HI\ tail is caused by a galactic wind, the latter should be highly asymmetric given that the tail is seen only on one side and also there are no signs of double peaks in the line profiles that could point to an expanding shell.
A second possibility is that of ram pressure stripping. This would justify the head-tail morphology of the HI in VV124 and also the smooth and small gradient in the tail.
The problem with this interpretation is the emptiness of the field around VV124, which forces us to assume that the ambient medium responsible for the stripping should be the intra-group medium associated with the Local Group. We can estimate the required density of the ambient medium in this way \citep{Grcevich09}: 

\begin{equation}
n_{\rm IGM} \sim \frac{ \sigma_{\rm HI}^2 \, n_{\rm HI} }{ 3\, v_{\rm LG}^2} \, {\rm cm^{-3}}
\end{equation}

where $\sigma_{\rm HI}$ is the measured \HI\ velocity dispersion, assuming that it is somehow representative of the potential (see below), $n_{\rm HI}$ is the gas volume density, $V_{\rm LG}$ is the velocity of the galaxy with respect to the LG medium, $V_{\rm LG} = 72.6 \,{\rm km\,s^{-1}}$ \citep[computed with Eq.~11 by][]{tully}.
Using the numbers relevant for VV124, we estimate that the density of the ambient medium should be $n_{\rm IGM} \sim 5 \times 10^{-4} \, {\rm cm^{-3}}$, which is probably too large at the distance of VV124 from the centre of the Local Group \citep[e.g.][]{ande}. For example, this value is larger than the density required to produce the Magellanic Stream (at mere 50~kpc from the MW) in the model by \citep{mastrop}. Hence, while both scenarios outlined above do not provide a fully satisfactory explanation for the \HI\ tail and gradient, the stellar wind hypothesis appears more likely, as a low-intensity recent star-formation episode is actually observed to occur near the center of VV124 (see Sect.~\ref{cmd} and J10).

It is worthwhile to note that the distribution of \HI\  does not appear to correlate with the wing-like substructures discussed in Sect.~\ref{dmap} above. Moreover, Fig.~\ref{velfie} shows that virtually all the detected \HI\  is enclosed within the $\mu_r=26.0$~mag/arcsec$^2$ optical isophote, having semi-major axis $\simeq 2\arcmin$. The $\mu_r=30.0$~mag/arcsec$^2$ optical isophote has semi-major axis $\simeq 4\arcmin$, hence the \HI\  contours do not exceed the optical body of the galaxy.

\begin{figure}
\centering
\includegraphics[width=\columnwidth]{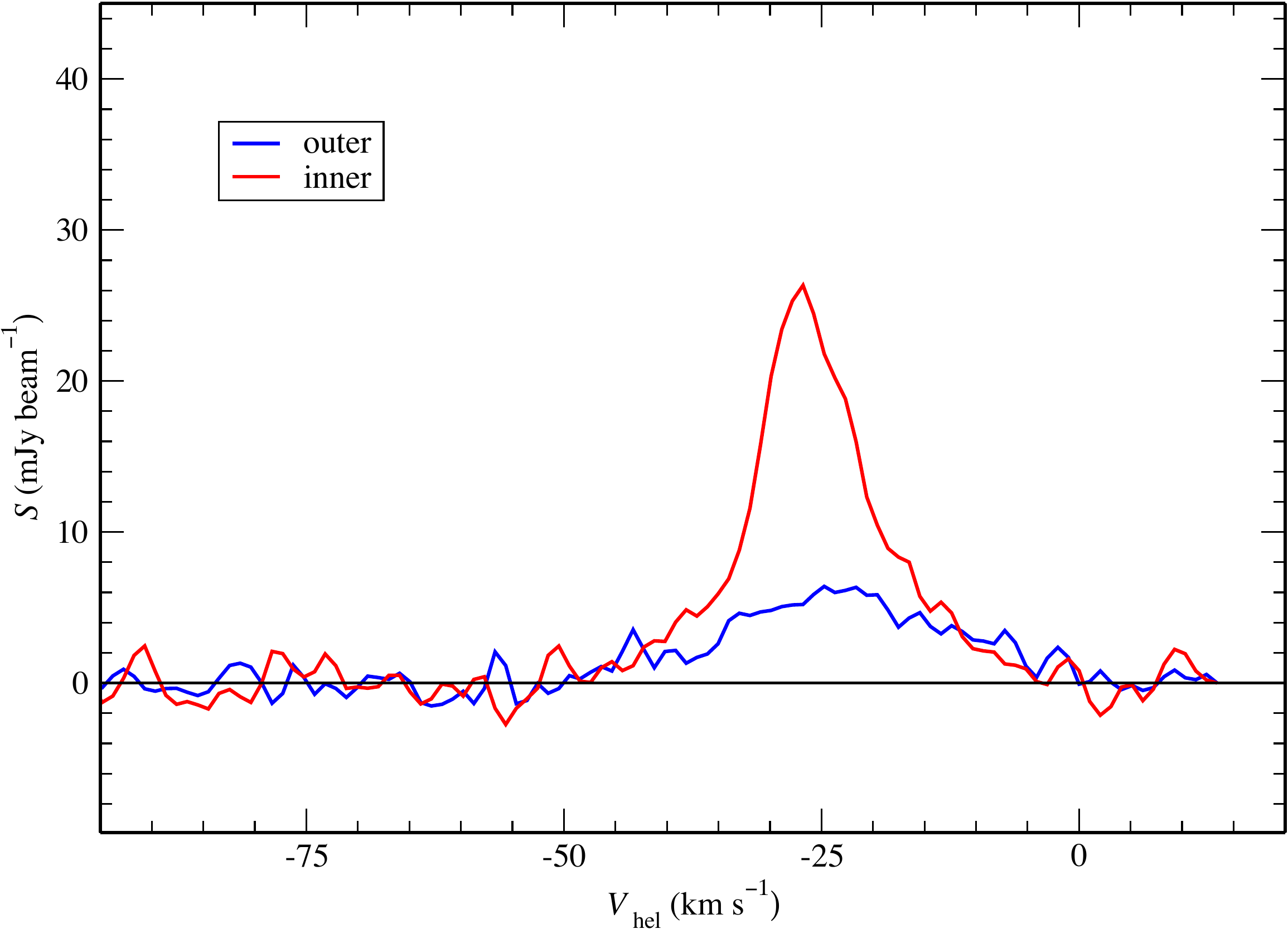}
\caption{Two \HI\  profiles showing the presence of a two-phase ISM in VV124. The profile taken over a 1 arcmin $\times$ 1 arcmin region  of the low column density on the eastern side of VV124 only shows a broad component, while the profile taken over the region where the column density is above  $2\times 10^{20}$ cm$^{-2}$ shows the presence of an additional narrow component}
\label{HIprof}
\end{figure}

The \HI\  spectra do indicate the presence of a two-phase interstellar medium. In the regions where the column density is below $2\times 10^{20}$ cm$^{-2}$, the spectra show a single component with a velocity dispersion of, on average, 11 km s$^{-1}$ (see Fig.~\ref{HIprof}). However, in the regions where the column density is above this value, the spectra show, in addition, a narrower component with a dispersion of about 4 km s$^{-1}$. This two-phase character of the ISM is seen in many dwarf galaxies \citep{Young96,Young97,Begum06,Ryan08}.

A rough estimate can be made of the dynamical mass of VV124 by assuming a spherical and isotropic \HI\  distribution. In this case,  the dynamical mass is given by $M_{\rm dyn} = 5 r_{\rm HI} \sigma^2/G$. For a velocity dispersion of 11 km s$^{-1}$ and a size of the \HI\  of $r_{\rm HI} = 475 $ pc this gives $M_{\rm dyn} = 6.6 \times 10^7\ M_\odot$. This would mean that $M_{\rm dyn}/M_{\rm baryon} \simeq 4$ and $\frac{M}{L_V}\simeq 8$. Such a baryon content is very typical for small isolated galaxies such as VV124 \citep{Begum06}. 

\subsection{Low Resolution Spectroscopy}
\label{LRS}

Triggered by the apparent mismatch between the very accurate systemic velocity obtained from \HI\  data ($V_{\rm h}\simeq -25$~km~s$^{-1}$) and the estimate from low-resolution ($R=\lambda/\Delta\lambda\simeq 1000$) optical spectroscopy by K08 ($V_{\rm h}\simeq -70$~km~s$^{-1}$), we applied for Director Discretionary Time (DDT) at the TNG, to obtain a new independent estimate of the radial velocity from optical spectra. We were awarded $\simeq 3$~hr to observe VV124 with the low-resolution spectrograph DoLoRes\footnote{\tt http://www.tng.iac.es/instruments/lrs/}.

   \begin{figure}
   \centering
   \includegraphics[width=\columnwidth]{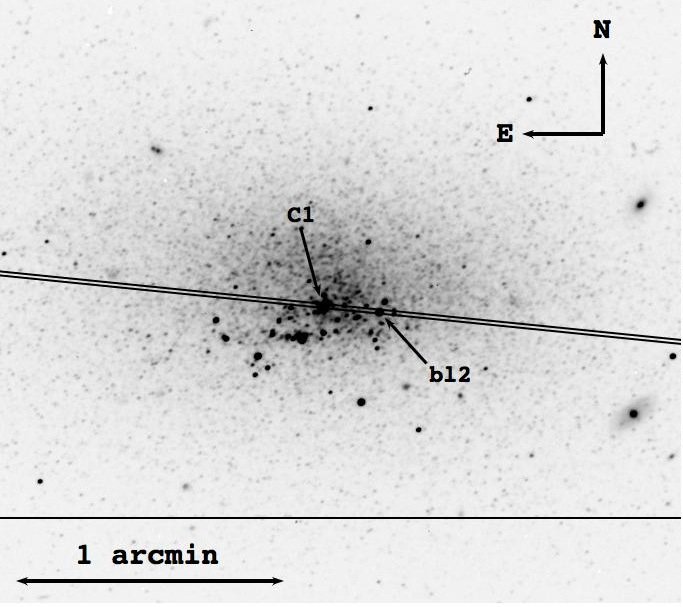}
     \caption{Slit position for our DoLoReS observations, superposed to a LBT $g$ image. The main targets are labeled.}
        \label{slit}
    \end{figure}


The camera is equipped with a $2048\times 2048$ px$^2$ E2V 4240 thinned, back-illuminated CCD.
The field of view is $8.6\arcmin \times 8.6\arcmin$ with a 0.252 arcsec/pix scale.
Observations have been acquired trough a long-slit $1\arcsec$ wide, using the VHRV holographic grism, which provides spectra in the range $4752\AA \le \lambda\le 6698\AA$, with spectral resolution $R=1527$ at $5725\AA$. The slit was positioned as shown in Fig.~\ref{slit}. The main targets of our observations were (a) the bright star bl2, already observed by K08, who classified it as spectral type F5Ia, and (b) the central source C1, which appears extended in our LBT images, hence it may be a (nuclear?) star cluster; C1 was not observed by K08. Both sources have a total magnitude $V\sim 18$, but C1 is expected to give a lower S/N spectrum because of its extended nature.
In addition, the slit crosses the whole galaxy nearly along its major-axis, thus spectra of the integrated light  can also be obtained.

\begin{figure}
   \centering
   \includegraphics[width=\columnwidth]{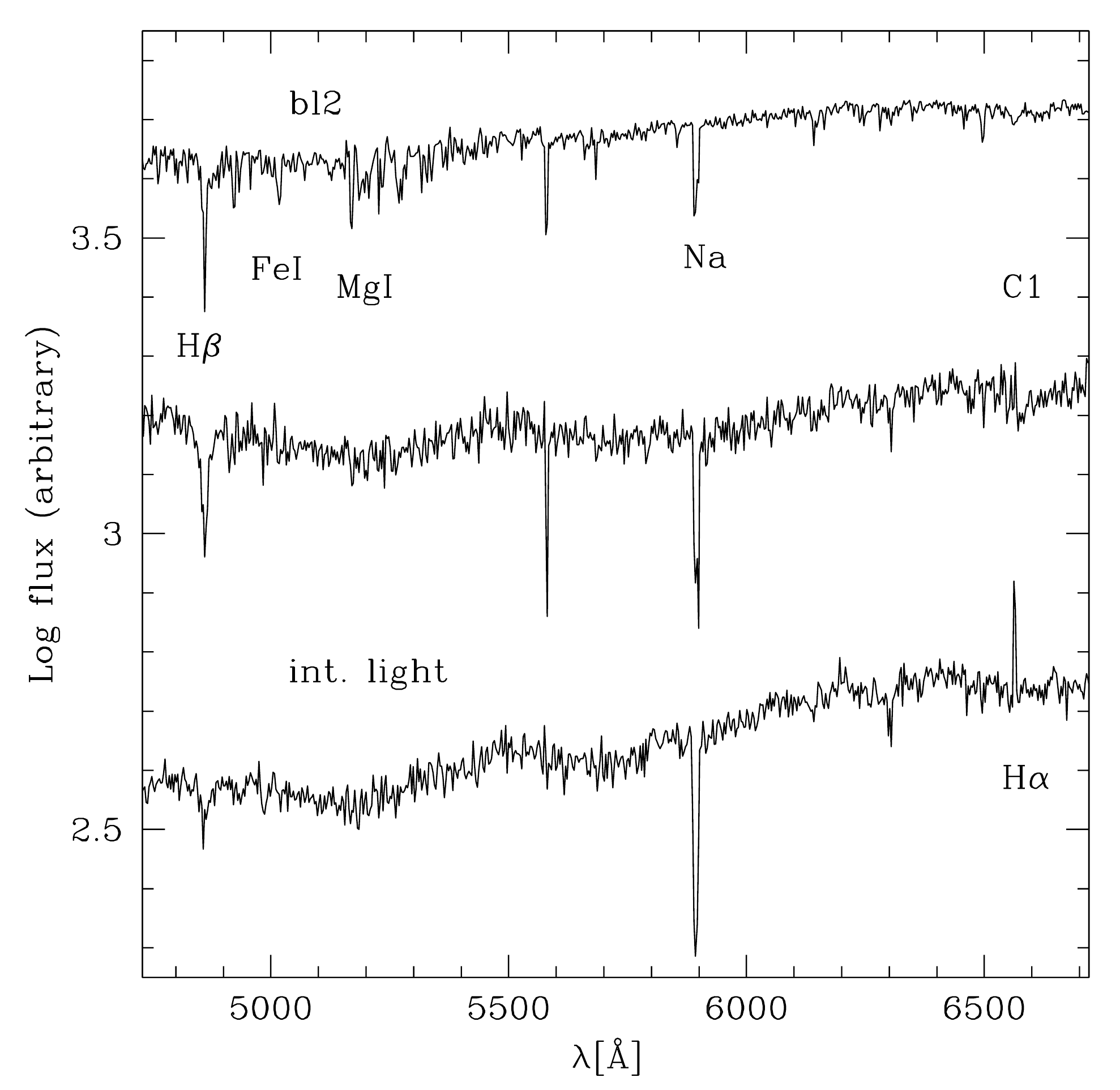}
     \caption{Spectra of the observed sources in logarithmic scale and arbitrary units.}
        \label{spettri}
    \end{figure}

The observations have been performed during the night of March 10, 2010.
We acquired four $t_{exp}=1800$~s spectra with the above-described set-up and slit position. In addition, Ar+Kr+Ne+Hg and He lamp spectra have been acquired immediately after each scientific exposure, to secure the most reliable wavelength calibration. Finally, proper bias and flat-field images were also obtained. The spectra of each source were corrected for bias and flat-field, extracted and wavelength-calibrated, and finally stacked together into a final spectrum using standard IRAF tasks \citep[see, for instance][and references therein]{silvr}. The co-added spectra of C1 and bl2 are shown in Fig.~\ref{spettri}, together with the integrated spectra of the region between the two sources. 

Several (relatively) strong absorption lines\footnote{A problem with the subtraction of a very strong sky line around
$5800\AA$ seriously affected the final shape of the Na absorption line at $5896\AA$ in all the spectra shown in Fig.~\ref{spettri}; for this reason this line has been always  excluded from the analysis.} are recognized and labeled in the spectrum of bl2, which is remarkably similar to that shown by T10 in the upper panel of their Fig.~6. On the other hand,  deep $H_{\beta}$ absorption is the only clearly recognized feature in the spectrum of C1, possibly suggesting a young age for this candidate cluster. Finally, the spectra of the integrated light, in addition to 
a weak $H_{\beta}$ absorption show an obvious $H_{\alpha}$ emission, indicating the presence of some hot gas, as also found by K08 and T10.

The heliocentric radial velocities of bl2 and C1 have been obtained by cross-correlation with a set of ten templates\footnote{The templates are high S/N spectra of M31 globular clusters with accurately known radial velocities, observed with the same set up \citep[][]{silvr}. As the typical integrated spectral type of GCs is F \citep{harris}, they are appropriate for the present application.}, using the IRAF {\tt fxcorr} task, following the procedure described in detail in \citet{silvr}. As our final value of $V_{\rm h}$ for each source we took the average from the ten semi-independent estimates. The standard deviation was adopted as the error. The estimates were repeated using the centroid of the strongest lines and the results were found to agree with those obtained from cross-correlation within the uncertainties. The velocity of the integrated light spectrum was obtained only from the centroid of the $H_{\alpha}$ emission line.
The zero point of our velocity scale was checked against the centroid of sky emission line and found to be accurate within $\simeq \pm 10$~km~s$^{-1}$ (note, however that the bluest accessible sky line was $5460.74\AA$ Hg; hence the stability of the wavelength calibration solution could not be checked in the bluest portion of the spectra).

\begin{table}
  \begin{center}
  \caption{Radial Velocities from DoLoRes spectra}\label{RV}
  \begin{tabular}{lcr}
    \hline
    Source & $V_{\rm h}$   & \\
           & [km~s$^{-1}$]      &     \\
\hline
bl2        & $-44\pm 18$  & cross correlation\\
C1         & $-86\pm 20$  & cross correlation\\
int. light & $-6 \pm 30$  &$H_{\alpha}$ emission\\
\hline
\end{tabular}
\end{center}
\end{table}

Before presenting our new velocity estimates, it is worth to recall the results of K08 and T10.
They found $V_{\rm h}=-90\pm 15$~km~s$^{-1}$ and $V_{\rm h}=-82\pm 15$~km~s$^{-1}$ for the two supergiants
bl1 and bl2, respectively, and $V_{\rm h}=-70\pm 15$~km~s$^{-1}$ for the overall integrated light (absorption lines). From the [OIII]$5007\AA$ emission in the integrated spectra of different regions they found  $V_{\rm h}=-71\pm 10$~km~s$^{-1}$ for their region $a$-slit2,
$V_{\rm h}=-47\pm 15$~km~s$^{-1}$ for region $b$-slit2, and $V_{\rm h}=-54\pm 15$~km~s$^{-1}$ for region $a$-slit1. Finally, they obtained two independent spectra of the \HII\ region they identified to the South-East of C1, finding $V_{\rm h}=-36$~km~s$^{-1}$ from slit1 and $V_{\rm h}=-55$~km~s$^{-1}$ from slit2. T10 conclude that, while all the above estimates are lower (sometimes significantly) than the mean systemic velocity derived from the \HI\  observations ($V_{\rm h}=-25\pm 5$~km~s$^{-1}$), they fall in the 0~km~s$^{-1}$ to $-50$~km~s$^{-1}$ range spanned by the overall \HI\  distribution associated to the galaxy. However, the inspection of the \HI\  velocity field shown in Fig.~\ref{velfie}, reveals that, in fact, the radial velocities of the neutral Hydrogen associated with VV124 range between $V_{\rm h}\simeq-20$~km~s$^{-1}$ and $V_{\rm h}\simeq-40$~km~s$^{-1}$.
Moreover, the sources having $V_{\rm h}\sim -70$ km~s$^{-1}$ from T10 optical photometry are projected on the side of the \HI\  distribution where the velocity of the gas is anywhere larger than $V_{\rm h}\simeq -27$ km~s$^{-1}$. Hence the correlation between the various optical velocities by T10 and the \HI\  velocity gradient is quite poor.

Our results, summarized in Tab.~\ref{RV}, do not add very much to the puzzling scenario presented by T10. The newly derived velocity of C1 matches very well the $V_{\rm h}\le -70$~km~s$^{-1}$ values that T10 find for bl1, bl2 and the integrated light. On the other hand the velocity we obtain from the integrated $H_{\alpha}$ emission is compatible within $1\sigma$ with the \HI\  mean velocity. bl2 is the most interesting case as (a) it is the source for which we got the highest S/N spectra and the richest one in terms of usable spectral lines, and (b) the target is in common with T10, thus providing a direct comparison between the two analyses. While the S/N of our spectra and the T10 one are very similar, the spectral resolution of our set-up is higher by a factor of $\sim 1.5$. Indeed, our estimate of $V_{\rm h}=-44\pm18$~km~s$^{-1}$ is at $\simeq 1\sigma$ from the \HI\  one, while the estimate by T10 is at $3.6\sigma$ from that value. 

 We conclude that the newly obtained $V_{\rm h}$ measures may help to reconcile optical and radio estimates of the systemic radial velocity of VV124, while still leaving reasons of concern (see Sect.\ref {disc}, for dicussion).

\section{Summary and Discussion}
\label{disc}

We have presented the results of deep wide-field photometry, optical low-resolution spectroscopy and \HI\  observations of the dwarf galaxy VV124=UGC4879, recently recognized as lying in the outer fringes of the LG. The main results of our analysis can be summarized as follows:

\begin{itemize}

\item While a sparse population of young stars (age$\la 500$~Myr) is observed in the inner region of the galaxy, the dwarf is dominated by an old population. There are indications of the presence of age/metallicity gradients, with older and more metal-poor stars being more prevalent at larger distances from the center of the galaxy.

\item We used a very clean detection of the RGB tip to obtain a new and more accurate distance estimate with respect to K08 and T10, $D=1.3\pm 0.1$~Mpc, in excellent agreement with the very recent estimate obtained by J10 from HST data. In spite of the larger distance, VV124 can be considered as a (present or future) member of the Local Group of galaxies, according to the criterion adopted by \citet[][see also the discussion and references on the method]{mateo}, and illustrated in Fig.~\ref{apex}, here. 

\item Independently of the actual membership to the LG, it is confirmed that VV124 is the most isolated nearby galaxy, very likely never disturbed by even a weak interaction with other dwarf or giant galaxies during its whole lifetime (see K08 and J10, for further discussion). For example, the relatively high recession velocity 
($V_g$=+100~km~s$^{-=1}$)  of the gas-poor isolated dwarf Tucana \citep{tuc} may indicate that it has been ejected to large distances by a three-body encounter, as envisaged by \citet{sales}.
Such a scenario cannot be invoked in the case of VV124, which has a much smaller galactocentric velocity 
\citep[$V_g$=+17~km~s$^{-=1}$; see also][for other scenarios for the transformation of dwarf galaxies and discussion]{donghia,kaza}.

   \begin{figure}
   \centering
   \includegraphics[width=\columnwidth]{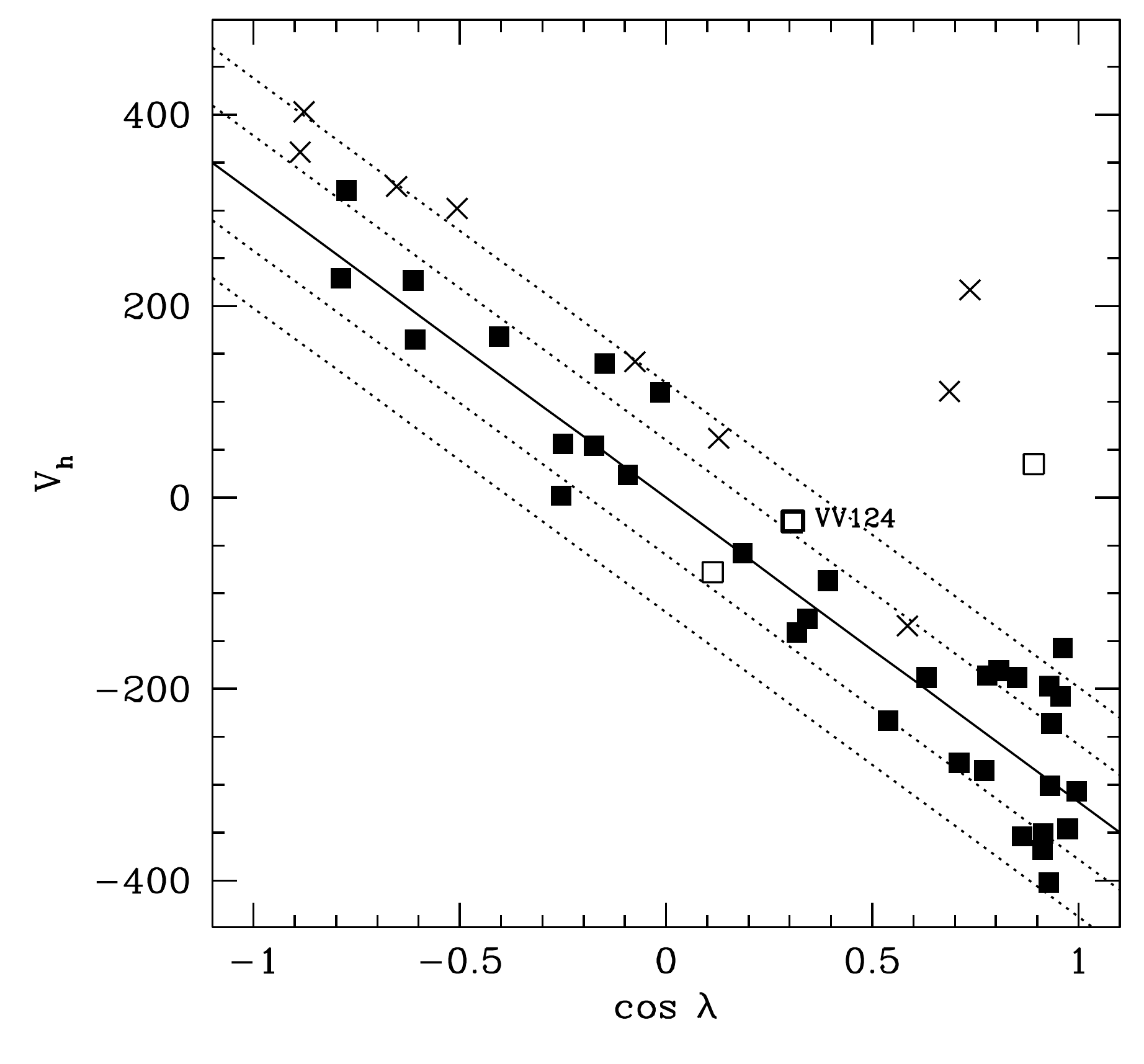}
     \caption{Heliocentric velocities versus the cosine of the angle between a galaxy and the apex of the Sun motion with respect to the center-of-mass of the LG \citep[see][and references therein]{mateo}, for galaxies within 2~Mpc from the barycenter of the LG ($D_{LG}<2$~Mpc) in the catalogue of \citet{tully}, plus VV124 (data from the present paper). 
The parameters of the solar motion are also taken from Tully et al., while the position of the barycenter is from \citet{vdb00}. Filled squares are galaxies with 
$D_{LG}\le 1.0$~Mpc, open squares have $1.0$~Mpc~$<D_{LG}\le 1.5$~Mpc, $\times$ symbols have $D_{LG}> 1.5$~Mpc. VV124 has $D_{LG}\simeq 1.3$~Mpc.
The continuous line is the locus of rest with respect to the center-of-mass of the LG, the dotted lines are at $\pm 1\sigma$ and $\pm 2\sigma$ from that locus \citep[$\sigma=60$~km~s$^{-1}$, from][and in good agreement with what found here from galaxies having $D_{LG}\le 1.3$~Mpc]{mateo}.}
        \label{apex}
    \end{figure}


\item The main visible body of VV124 is remarkably elliptical ($\epsilon\simeq 0.44$), regular and smooth, if the asymmetrically distributed minor young component is ignored. We were able to follow the SB profile of the galaxy out to a distance never reached before, $r_{\epsilon}\sim 1.9$~kpc, demonstrating that the galaxy is much more extended than previously believed. 

\item In addition, we obtained surface density maps of RGB stars revealing the presence of two relatively thin symmetric wings, emanating from the western and eastern edges of the inner elliptical body and aligned along its major axis, extending out to $\simeq 3$~kpc from the center of the galaxy. These low SB features ($\mu_r\ga 30$mag/arcsec$^2$) have no counterpart in other known galaxies of the LG (but see below).

\item \HI\  emission was detected for the first time by our high-sensitivity WSRT observations. The \HI\  total mass ($\sim 1\times 10^6~M_{\sun}$) is small with respect to the overall stellar luminosity (mass), making VV124  more gas-poor than typical isolated dwarfs. The density peak of the gas is slightly offsett with respect of the optical center (by $\sim 400$~pc) and the distribution shows a tail in the SE direction, also corresponding with the region with the most negative velocity.
This may be indicative of some outflow/inflow process possibly tied to the most recent star formation event (see J10), since the youngest stars are also preferentially found to the SE of the galaxy center. 

\item It is important to recall that {\em the observed \HI\  contours do not extend beyond the optical body of the galaxy as traced by RGB star counts}. Also, the \HI\  distribution do not show any obvious correlation with the surface density stellar wings described above.

\item The velocity field of the \HI\  does not show any sign of overall rotation over the main optical body of the galaxy, while a velocity gradient is observed in the tail-like structure in the SE side. The velocity gradient may indicate either infall or outflow of \HI. The projected size of the tail and the projected difference in velocity with the  main body give a timescale of several times $10^7$ yr. This timescale is very uncertain (due to projection effects), but could be connected to some elevated activity in the galaxy some time ago (see J10). It could be an outflow due to the most recent episode of star formation, or the remnant of some \HI\  inflow which triggered the star formation.

\item The gas associated with VV124 lies in the velocity range -20~km~s$^{-1} \ge V_{\rm h} \ge$~-40~km~s$^{-1}$, with a mean systemic velocity of $V_{\rm h}=-25\pm 4$~km~s$^{-1}$. The interstellar medium is found in two phases, a narrower component associated with the inner regions (at higher column density), and broader component found over the whole body of the \HI\  distribution, having a typical dispersion of 11~km~s$^{-1}$. 

\item The correlation between the velocity of various sources associated to VV124, as derived from optical spectra by us, K08 and T10, and the \HI\  velocity field is quite poor. 
While the observed differences can be accommodated within the uncertainties, it is a bit disturbing that the large majority of optical estimates lie at $V_{\rm h}<-40$~km~s$^{-1}$, beyond the lower side of the range of \HI\  velocities, in particular if one considers that they are now taken from two independent sources, i.e. T10 and this work. This  means that one should consider the possibility that there is a real difference between the systemic velocities of the stars and of the gas, due to the hypothesized gas flow suggested by the asymmetric structure of the \HI.
Coming Keck-DEIMOS observations of RGB stars in the galaxy will hopefully settle this issue.

\item Fig.~\ref{scale} shows that the structural parameters of VV124 are consistent with the scaling laws obeyed by dSphs and dIrrs galaxies \citep[see][and references therein]{korme,tht}. It is interesting to note that it lies at the upper envelope of the $M_V - \mu_V$ relation, i.e. it has the brightest SB for its total luminosity, and in an intermediate position between dIrrs and dSphs in  $M_V - r_h$ relation. According to \citet{pena}, the evolution in a relatively strong tidal field would have led to a larger scale radius, a lower luminosity and a lower surface density, thus driving VV124 toward the loci of ''genuine'' dSphs, in these planes.
This is in qualitative agreement with the possibility that VV124 may be a precursor of modern dSphs that did not enter in the interaction-driven evolutionary path that produced the latter class of dwarf galaxies.


\end{itemize}

   \begin{figure}
   \centering
   \includegraphics[width=\columnwidth]{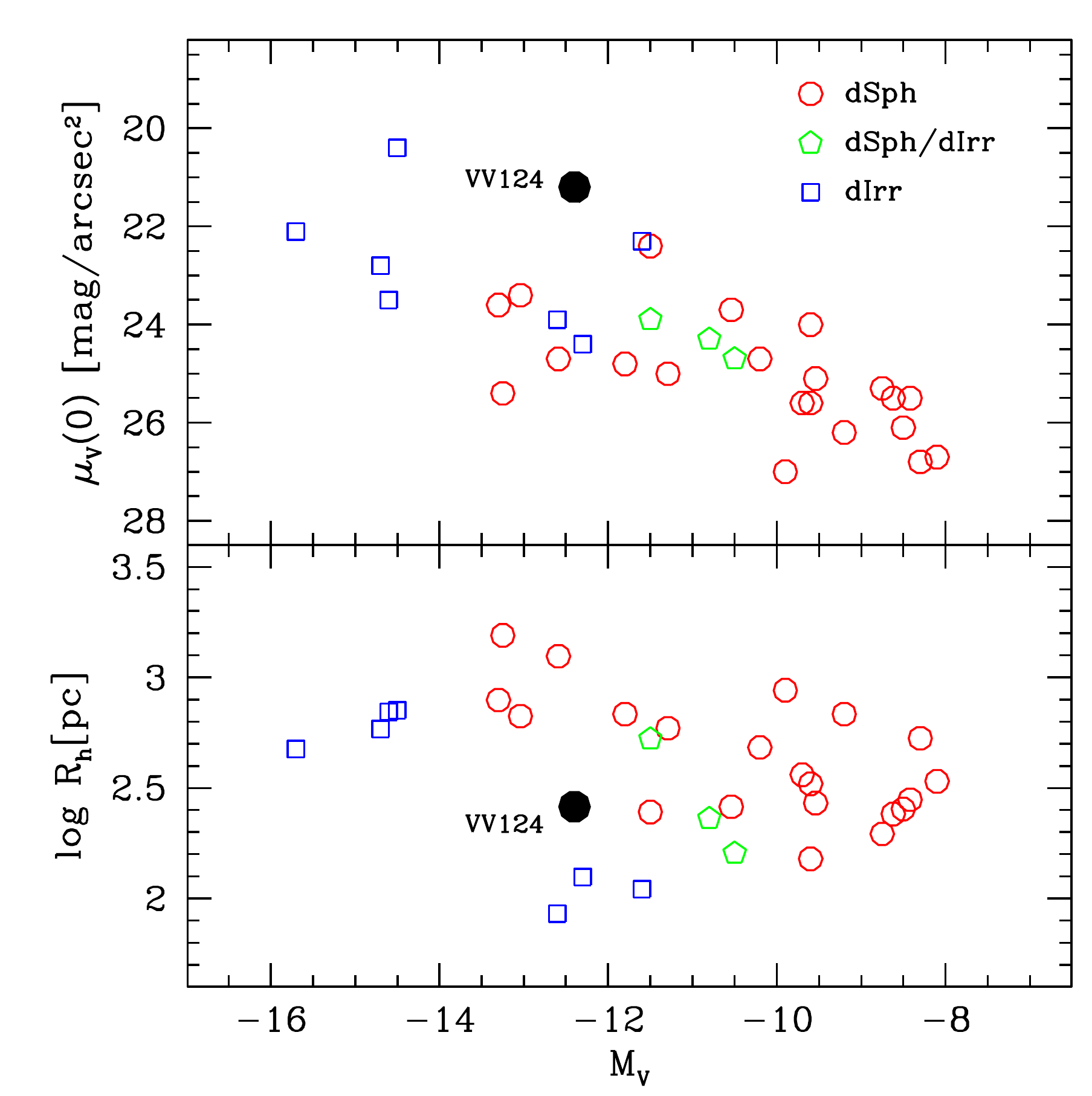}
     \caption{VV124 (black filled circle) compared to other LG dwarf galaxies brighter than $M_V=-7.0$ in the $M_V$ vs. logarithm of the half-light radius (lower panel) and $M_V$ vs. central surface brightness (upper panel) planes. Different symbols are used for different morphological types, according to the classification criteria by \citet{mateo}. Absolute magnitudes and half-light radii are taken from from \citet{walk10} and from \citet{kalirai}; when lacking in these sources they have been drawn from \citet{mateo}.
Surface brightness estimates have been taken from \citet{mateo}, \citet{mccon}, \citet{mart09}, \citet{zuck}, \citet{mike}, and \citet{whiting}, with this priority.
All the values of surface brightness are corrected for extinction.
     }
        \label{scale}
    \end{figure}


Probably the most intriguing features of VV124 revealed by the present study 
are the low SB stellar wings; in particular if they are interpreted as an ancient stellar disk 
seen edge-on and considering the extreme isolation in which the galaxy 
evolved. As mentioned in the introduction, 
the idea that present-day dwarf spheroidals originated from dIrr galaxies that 
were deprived from their gas at early times, as a result of tides
and/or ram pressure stripping within the potential of a much larger galaxy or of a group/cluster, 
is not new \citep[see][and references therein]{mateo,tht}. The model recently developed by \citet{lucio1,lucionat} 
explicitly postulates dwarf disk galaxies as the progenitors of dSphs 
\cite[see][for a recent and thorough discussion]{kaza}. 
Indeed, observational evidence of the existence of the expected intermediate stages of such
transformation process is beginning to emerge, 
from study of various environments \citep[see, for example, the disky dEs identified in the Virgo cluster by][]{lisker}. 
In the context of the LG, \citet{n205ivo} have recently reviewed the evidence for the presence of the relics of a disk 
in NGC~205, suggesting that the process of transformation of the disk into a spheroid is currently ongoing in 
this satellite of M31. Two independent theoretical studies have shown that some relevant observational properties of 
the disrupting Sagittarius dSph and of the associated tidal stream can be more easily explained if a disk galaxy is 
adopted as the progenitor of Sgr \citep{pena_sgr,lokas_sgr}. NGC~205 and Sgr may simply represent two  
different stages on a similar evolutionary path driving the transformation of a similar low luminosity
disky progenitor into a dSph via the interaction with the main
galaxy. In this framework, VV124 would be akin to such disky progenitors but {\em never} entered 
the transformation path. Rather, it evolved passively in isolation, thus preserving its disk intact 
until the present day.  Assuming we are looking at a rotationally supported stellar system, the disk
of VV124 would be relatively thick, but this is normally seen in the faintest dIrrs in the Local Group
and nearby clusters (Sanchez-Janssen et al. 2010). The thickness of the disk does not reflect
environmental effects, but is rather the result of
pressure support from either internal feedback from star formation and/or the cosmic ultraviolet ionizing
background becoming increasingly more important for the energy balance towards increasingly lower masses. Kaufmann, Wheeler
\& Bullock (2007) have shown how, for galaxies having circular velocities $< 30$ km/s 
hosted in halos with typical spin parameters ($\lambda 
< 0.05$), an effective temperature (i.e. thermal + turbulent) of the ISM of a few times $10^4$ K, or equivalently an ISM velocity
dispersion of $\sim 10-12$ km/s, is sufficient to produce a substantially thick disk (with major-to-minor axis
ratio in the range $0.2-0.4$) as an equilibrium configuration. This is because in such systems the gas velocity
dispersion is already close to the virial temperature (for circular velocities below $30$ km/s the virial temperature
is $< 5 \times 10^4$ K), which forces the gas to acquire a high pressure scale height at equilibrium. Stars in such a system form out of a pressure supported turbulent gas disk and, being collisionless, have no way to 
dissipate such motions later on. Again, in the Local Group, at the lowest luminosity end of dIrrs, there are a number
of examples of dwarfs with \HI\  velocity dispersions around $10$ km/s (SagDig, Leo A and GR8 being some of these),
and whose low rotation velocities ($< 15$ km/s) implies a halo circular velocity well below $30$ km/s, consistent
with the picture just outlined.

Another very important issue is to understand how the galaxy became gas-poor in absence of stripping mechanisms 
due to the interaction with other galaxies. Photo-evaporation of the gas after 
re-ionization \citep{ioniz} is a possibility certainly worth further investigation
\citep[see, e.g.][]{susa}.
Supernovae feedback might also play a role; recent cosmological simulations of the formation of a gas rich dwarf
that are finally able to produce a realistic exponential disk with no bulge (Governato, Brook, Mayer et al. 2010)
show that outflows at high redshift ($z > 1$) can remove more than $2/3$ of the baryons even in dwarfs with
circular velocities exceeding $V_{circ}$ = 30 km/s. Finally, given the fragility of such a low mass galaxy, we cannot
exclude that the interaction with intergalactic gas clouds in the Local Group could have caused stripping of at least a fraction of an already diffuse,
loosely bound interstellar medium.

In this context it is very interesting to recall that also the Dark Matter (DM) halo  expected to embed VV124 should be virtually untouched since the epoch of its collapse: the kinematics of the stars in the main body and in the wings of the galaxy should probe the mass profile of this pristine halo out to a remarkably large radius ($\sim 3$~kpc), possibly opening a crucial window on the initial conditions of DM halos of this low-mass scale. 


\begin{acknowledgements}

We are grateful to Judith Cohen and Monica Tosi for a critical reading of the manuscript.

Based on data acquired using the Large Binocular Telescope (LBT). The LBT is an international collaboration among institutions in the United States, Italy, and Germany. LBT Corporation partners are The University of Arizona on behalf of the Arizona university system; Istituto Nazionale di Astrofisica, Italy; LBT Beteiligungsgesellschaft, Germany, representing the Max-Planck Society, the Astrophysical Institute Potsdam, and Heidelberg University; The Ohio State University; and The Research Corporation, on behalf of The University of Notre Dame, University of Minnesota and University of Virginia. 

Based on data acquired at the Westerbork Synthesis Radio Telescope (WSRT)
The WSRT is operated by the Netherlands Institute foir Radio Astronomy (ASTRON) with the support from the Netherlands Foundation for Scientific Research (NWO). 

Based on observations made with the Italian Telescopio Nazionale Galileo (TNG) operated on the island of La Palma by the Fundaci—n Galileo Galilei of the INAF (Istituto Nazionale di Astrofisica) at the Spanish Observatorio del Roque de los Muchachos of the Instituto de Astrofisica de Canarias. We are especially grateful to the Director of TNG E.~Molinari and the TNG DDT TAC for giving us the possibility to get spectra of VV124 using Director Discretionary Time, and W.~Boschin for the assistance in the preparation of the observing blocks.

This research make use of SDSS data. 
Funding for the SDSS and SDSS-II has been provided by the Alfred P. Sloan Foundation, the 
Participating Institutions, the National Science Foundation, the U.S. Department of Energy, the National Aeronautics and Space Administration, the Japanese Monbukagakusho, the Max Planck Society, and the Higher Education Funding Council for England. The SDSS Web Site is http:\/\/www.sdss.org\/.
The SDSS is managed by the Astrophysical Research Consortium for the Participating Institutions. The Participating Institutions are the American Museum of Natural History, Astrophysical Institute Potsdam, University of Basel, University of Cambridge, Case Western Reserve University, University of Chicago, Drexel University, Fermilab, the Institute for Advanced Study, the Japan Participation Group, Johns Hopkins University, the Joint Institute for Nuclear Astrophysics, the Kavli Institute for Particle Astrophysics and Cosmology, the Korean Scientist Group, the Chinese Academy of Sciences (LAMOST), Los Alamos National Laboratory, the Max-Planck-Institute for Astronomy (MPIA), the Max-Planck-Institute for Astrophysics (MPA), New Mexico State University, Ohio State University, University of Pittsburgh, University of Portsmouth, Princeton University, the United States Naval Observatory, and the University of Washington.

\end{acknowledgements}

\bibliographystyle{apj}



\end{document}